\documentstyle[aps,epsf,floats,twocolumn]{revtex}

\def\figwidth{0.4\columnwidth}
\def\figwidtha{0.8\columnwidth}
\def\figwidthb{0.4\columnwidth}
\def\figwidthc{0.4\columnwidth}
\def\postscript#1{\begin{center}\leavevmode \hbox{\epsfxsize=\figwidth\epsfbox{#1}}\end{center}}

\begin{document}
\markboth{cond-mat/9612106 -- to appear in PHYSICAL REVIEW B1 01Mar97}{cond-mat/9612106 -- to appear in PHYSICAL REVIEW B1 01Mar97}
%\draft

\title{Triplanar Model for the Gap and Penetration Depth in YBCO}
\author{C. O'Donovan\cite{me} and J. P. Carbotte}
\address{Department of Physics \& Astronomy, McMaster University, \\
 Hamilton, Ontario, Canada L8S 4M1}
\date{October 10, 1996}

% This is the secret. The closing ``]'' on the line below...
\twocolumn[\hsize\textwidth\columnwidth\hsize\csname@twocolumnfalse\endcsname
\maketitle
\begin{abstract}
YBaCuO$_7$ is a trilayer material with a unit cell consisting of a
CuO$_2$ bilayer with a CuO plane of chains in between.  Starting with
a model of isolated planes coupled through a transverse matrix
element, we consider the possibility of intra as well as interplane
pairing within a nearly antiferromagnetic Fermi liquid model.
Solutions of a set of three coupled {\sc bcs} equations for the gap
exhibit orthorhombic symmetry with $s$- as well as $d$-wave
contributions.  The temperature dependence and $a$-$b$ in plane
anisotropy of the resulting penetration depth is discussed and
compared with experiment.
\\{\tt preprint: cond-mat/9612106}
\end{abstract}
% ...is the rest of the secret.
]

\narrowtext
\section{Introduction}

     Within the unit cell of YBa$_2$Cu$_3$O$_7$, there are two CuO$_2$
planes and a single set of CuO chains oriented along the $b$-axis.
The chains break the tetragonal symmetry of the system and make it
orthorhombic.  The role of the chains is not yet fully understood but
it is clear the they lead to a large in-plane anisotropy in various
properties such as the value of the zero temperature penetration depth
and the {\sc dc} conductivity
etc.\cite{basov,zhang,bonn,friedmann,iye,gagnon,takenaka} For example,
in optimally doped YBa$_2$Cu$_3$O$_{6.95}$, the penetration depth
along the $a$-axis at $T=0$ is $\lambda_a(0)\sim 1600$\AA\ while that
along the $b$-axis is $\lambda_b(0)\sim1030$\AA.\cite{zhang} At 300 K
the ratio of the {\sc dc} resistivities $\rho_b/\rho_a$ is larger than
a factor of two.  These large anisotropies would not exist in a
purely tetragonal system.  The role of the chains in YBCO has been
investigated before using a two plane model\cite{atkinson} as well as
a plane-chain model.\cite{atkinson2,odonovan10} While these
calculations have given us considerable insight, such a model is not
fully realistic.  Here we want to explicitly introduce the trilayer
nature of the system.  We start with a pair of isolated CuO$_2$ planes
and a set of chains (CuO plane).  They are assumed to be coupled
through a transverse matrix element, $t_1$, between the chain and each
of the planes and $t_2$ between the two planes.  A set of three
coupled {\sc bcs} equations are used to describe the pairing with
intra- as well as interband coupling.  For the pairing potential, we
use the form of the magnetic susceptibility introduced in the
phenomenological nearly antiferromagnetic Fermi liquid model of
Monien, Millis and Pines\cite{mmp} ({\sc mmp}) which we multiply by a
dimensionless parameter $g_{\alpha\beta}$ which allows us to vary at
will the strength of the diagonal (in plane) and off diagonal
(interplane) couplings.  Here the indices $\alpha\beta$ range over
the three subbands.  Numerical solutions of these equations using a
fast Fourier transform technique give solutions which are an admixture
of $d$, $s_\circ$ and $s$ symmetry.  For the tetragonal crystal point
group, these three symmetries would belong to two different
irreducible representations and would, in general, not mix.  But
because of the existence of the chains, the system possesses reduced
symmetry although, in most cases considered, nodes remain on the Fermi
surface and the linear low temperature variation of the penetration
depth remains.  In section II, we give formalism.  Solutions of the
gap equations are discussed in section III along with results for the
penetration depth.  A brief conclusion is included in section IV.

\section{Formalism}

     Consider a three plane per unit cell system with energies
$\varepsilon_{{\bf k},\alpha}$.  Here $\alpha =1,2,3$ is an index that
enumerates the three planes; two CuO$_2$ which are identical to each
other and one CuO plane consisting of chains oriented along the
b-direction.  The momentum index ${\bf k}$ is restricted to the two
dimensional CuO$_2$ Brillouin zone which has tetragonal symmetry.  A
tight binding model will be used to describe these isolated planes of
the form:\cite{odonovan4,odonovan3,odonovan1}
\begin{eqnarray}
\label{disp.eq}
\varepsilon_{{\bf k},\alpha}=
&-&2t_\alpha\left[(1+\epsilon_\alpha)\cos(k_x)+(1-\epsilon_\alpha)\cos(k_y)\right.\nonumber\\
&-&\left.2B_\alpha\cos(k_x)\cos(k_y)-(2-2B_\alpha-\mu_\alpha)\right],
\end{eqnarray}
\noindent
where $t_\alpha$ is the first neighbour hopping parameter,
$\epsilon_\alpha$ an orthorhombic distortion which can be used to
describe the chains, $B_\alpha$ is the second neighbour hopping in
units of $t_\alpha$, the $k_i$ are in units of reciprocal lattice
vectors and $\mu_\alpha$ is related to the chemical potential and
determines the filling.  Here it will be treated as a parameter.  In
what follows, the $t_\alpha$'s will be taken to be the same for
simplicity and denoted by $t_\circ$.  The three dimensional nature of
the system is introduced through transverse matrix elements, $t_1$,
between the CuO$_2$ planes and the CuO chains and $t_2$ directly
between the CuO$_2$ bilayer.  This changes the energies
$\varepsilon_{{\bf k},\alpha}$ to band energies
$\tilde{\varepsilon}_{{\bf k},\alpha}$ which we then use in the {\sc
bcs} gap equations assuming singlet pairing between equal and opposite
momentum pairs.  Including intraband (on diagonal) as well as
interband (off diagonal) pairing through the 2-body potential $V_{{\bf
k},{\bf k}^\prime,\alpha\beta}$ which describes scattering from ${\bf
k}$ to ${\bf k}^\prime$ with $\alpha$ and $\beta$ band indices the gap
in the $\alpha$$^{\rm th}$ band $\mit\Delta_{{\bf k},\alpha}$
satisfies the {\sc bcs} equations:\cite{odonovan10}
\begin{eqnarray}
\label{bcs.eq}
\mit\Delta_{{\bf k},\alpha}&=& \frac{1}{\Omega}\sum_{{\bf k}^\prime,\beta}{
V_{{\bf k},{\bf k}^\prime,\alpha\beta}\frac{\mit\Delta_{{\bf k}^\prime,\alpha}}{2E_{{\bf k}^\prime,\alpha}}
\tanh\left(\frac{E_{{\bf k}^\prime,\alpha}}{2k_{\rm B}T}\right)},
\end{eqnarray}

In equation (\ref{bcs.eq}), $\Omega$ is the crystal volume, $k_{\rm
B}$ Boltzmann's constant, $T$ temperature and
\begin{eqnarray}
\label{eigen.eq}
E_{{\bf k},\alpha}=\pm\sqrt{
\tilde{\varepsilon}_{{\bf k},\alpha}^2+\mit\Delta_{{\bf k},\alpha}^2}
\end{eqnarray}
are the quasiparticle energies of the $\alpha$$^{\rm th}$ band in the
superconducting state. The single particle normal state band energies
for a two layer model come from the Hamiltonian:
\begin{equation}
H = 
\left[\begin{array}{cc}
\varepsilon_1 & t_1 \\
t_1 & \varepsilon_2 
\end{array}\right]
\label{hamiltonian2.eq}
\end{equation}
with eigenvalues:
\begin{equation}
\tilde{\varepsilon}_{1, 2} = \left(\frac{\varepsilon_1+\varepsilon_2}{2}\right)
\pm \sqrt{\left(\frac{\varepsilon_1-\varepsilon_2}{2}\right)^2+t_1^2},
\end{equation}
and for the three layer model come from the Hamiltonian:
\begin{equation}
H = 
\left[\begin{array}{ccc}
\varepsilon_1 & t_1 & t_2\\
t_1 & \varepsilon_2 & t_1\\
t_2 & t_1 & \varepsilon_1
\end{array}\right]
\label{hamiltonian.eq}
\end{equation}
\noindent
with eigenvalues:
\begin{eqnarray}
\tilde{\varepsilon}_1 =&& \varepsilon_1-t_2
\nonumber\\ 
%&&{\rm (even-plane like)}\nonumber\\
%\label{params2.eq}
\tilde{\varepsilon}_2 = &&
\frac{\varepsilon_1+\varepsilon_2+t_2}{2}\nonumber\\
&&+ \frac{1}{2}\sqrt{(\varepsilon_1-\varepsilon_2)^2+2t_2(\varepsilon_1+\varepsilon_2+t_2)
+8t_1^2}
\nonumber
%\\&&{\rm (chain like)}\nonumber
%\label{params2.eq}
\end{eqnarray}
\noindent
and
\noindent
\begin{eqnarray}
\tilde{\varepsilon}_3 =&& 
\frac{\varepsilon_1+\varepsilon_2+t_2}{2}\nonumber\\
&&- \frac{1}{2}\sqrt{(\varepsilon_1-\varepsilon_2)^2+2t_2(\varepsilon_1+\varepsilon_2+t_2)
+8t_1^2}\nonumber
%\label{params2.eq}
\end{eqnarray}
\noindent
which are respectively the solid, dashed and dotted curves in Fig.\ \ref{fig5}.

     The set of three coupled equations (\ref{bcs.eq}) for the gap's
$\mit\Delta_{{\bf k},\alpha}$ in the $\alpha$$^{\rm th}$ band
($\alpha=1,2,3$) are explicit once the pairing potentials $V_{{\bf
k},{\bf k}^\prime,\alpha\beta}$ are specified.  Different models could
be used.  Here for simplicity and to be specific, we will assume that
the superconducting condensation proceeds through the spin
susceptibility $\chi_{{\bf k},{\bf k}^\prime}$ which for the nearly
antiferromagnetic Fermi liquid is peaked at the commensurate wave
vector $(\pi,\pi)$ at the corner of the Brillouin zone.  So as not to
introduce new parameters, here we will use the form of $\chi_{{\bf
k},{\bf k}^\prime}$ introduced through a fit to {\sc nmr} data by
Millis, Monien and Pines ({\sc mmp}).\cite{mmp} Their susceptibility and
consequent pairing potential has the form:
\begin{eqnarray}
V_{{\bf k},{\bf k}^\prime,\alpha\beta}=g_{\alpha\beta}
\frac{-t_\circ}{1+\xi_\circ^2|{\bf k}-{\bf k}^\prime-{\bf Q}|^2},
\label{mmp.eq}
\end{eqnarray}
\noindent
where the scale on $V_{{\bf k},{\bf k}^\prime,\alpha\beta}$ has been
set by $t_\circ$ for convenience and $g_{\alpha\beta}$ are
dimensionless parameters that determine the overall size of the
pairing as well as the relative amount of diagonal ($\alpha=\beta$)
and off diagonal ($\alpha\not=\beta$) coupling desired.  In all our
calculations, the relative sizes of the $g_{\alpha\beta}$ will be
chosen at will with absolute value adjusted to get a critical
temperature of 100K typical for the oxide superconductors.  The other
parameters in (\ref{mmp.eq}) are also specified and fixed at values
given by {\sc mmp}.  $\xi_\circ=2.5$\AA\ is the magnetic coherence
length and ${\bf Q}$ the commensurate wave vector $(\pi,\pi)$ in the 2
dimensional CuO$_2$ Brillouin zone.

     In writing down the set of equations (\ref{bcs.eq}), we have
assumed that the coherence length transverse to the planes is short so
that the gap can have different values in the different bands
$\tilde{\varepsilon}_{{\bf k},\alpha}$.  The model still allows for a
single gap, applicable to all three planes, to exist.  The appropriate
limit is $g_{\alpha\beta}$ independent of $\alpha$ and $\beta$ in
which case $\mit\Delta_{{\bf k},\alpha}$ on the left hand side of
(\ref{bcs.eq}) is clearly independent of $\alpha$ and a single gap
parameter describes the entire system.

     Numerical solution of the set of equations (\ref{bcs.eq}) using
fast Fourier transforms makes no assumption on the symmetry of the
resulting gap parameters $\mit\Delta_{{\bf k},\alpha}$ as a function
of momentum ${\bf k}$ in the 2 dimensional Brillouin zone.  In our
work, we find that these solutions consist of a mixture of
$d_{x^2-y^2}$, $s_\circ$ and $s_{x^2+y^2}$ symmetries with higher
harmonics as well as the more familiar lowest ones which are
\begin{mathletters}
\begin{eqnarray}
s_\circ \sim \eta_{\bf k}^{(s_\circ)}&\equiv& 1 \label{so.eq} \\
s_{x^2+y^2} \sim \eta_{\bf k}^{(s_x)}&\equiv& \cos(k_x)+\cos(k_y)\label{sx.eq} \\
d_{x^2-y^2} \sim \eta_{\bf k}^{(d)}&\equiv& \cos(k_x)-\cos(k_y).\label{d.eq}
\end{eqnarray}
\end{mathletters}

We are now ready to present results including the temperature
dependence of the penetration depth $\lambda(T)$ for various model
parameters.  To obtain the penetration depths, we have adopted the
following approximate procedure.  The lowest harmonics represented by
(\ref{so.eq}) to (\ref{d.eq}) are projected out of our numerical
solutions and these are used in all further calculations.  This
neglects higher harmonic contributions but is an adequate procedure
since we are mainly interested in qualitative results and trends.  In
any case, our model for the pairing interaction (\ref{mmp.eq}) is
simplified and is not believed to be accurate.  We use a {\sc bcs}
approach while it is necessary to go to an Eliashberg formulation to
get quantitative results.  This goes beyond the scope of this work.

     For completeness, the approximate formula for the penetration
depth employed here is:
\begin{eqnarray}
\label{pd2.eq}
\lambda_{ij}^{-2}=&&\frac{4\pi e^2}{c^2}\frac{1}{\Omega}\sum_{{\bf k},\alpha}
v_{\alpha,i}v_{\alpha,j}
\left(
\frac{\partial f(\tilde{\varepsilon}_{{\bf k},\alpha})}{\partial \tilde{\varepsilon}_{{\bf k},\alpha}}
-\frac{\partial f(E_{{\bf k},\alpha})}{\partial E_{{\bf k},\alpha}}
\right)
\end{eqnarray}
\noindent
where the constant $c$ is the speed of light, $v_{\alpha,i}$ is the $i$$^{\rm th}$
component of the electron velocity in the $\alpha$$^{\rm th}$ band
\begin{eqnarray}
v_{\alpha,i}\equiv
\frac{1}{\hbar}
\frac{\partial \tilde{\varepsilon}_{{\bf k},\alpha}}{\partial k_i}
\end{eqnarray}
\noindent
and $f(x)$ is the Fermi-Dirac distribution given by
\begin{eqnarray}
f(x)\equiv\left(e^{\tilde{\varepsilon}_{{\bf k},\alpha}/k_{\rm B}T}+1\right)^{-1}
\end{eqnarray}
\noindent
We are now in a position to present results.

\section{Results}

     Before presenting results for the trilayer model, it is
instructive to consider a bilayer consisting of a CuO$_2$ plane with
tetragonal symmetry and a set of CuO chains directed along the $b$
axis which has lower orthorhombic symmetry.  This difference in
symmetry has a drastic effect on the solution of the {\sc bcs} gap
equation.  To illustrate this, we take a model band structure:
\begin{equation}
\{t_\circ, \epsilon_\alpha, B_\alpha, \mu_\alpha\} 
=
\left\{
\begin{array}{rrrr}
50.0 & 0.0 & 0.45 & 0.51 \\
50.0 & 1.0 & 0.00 & 1.2
\end{array}
\right\}
\end{equation}
\noindent
with $t_\circ$ in meV and the other parameters are dimensionless.  We
also consider the pairing potential to be of the {\sc mmp} type and of
equal magnitude in both chains and plane.  So as to have a single
critical temperature, we also include a smaller off diagonal pairing
$g_{\alpha\beta}\not = 0$ for $\alpha\not =
\beta$.  Specifically, we take:
\begin{equation}
g_{\alpha\beta}=\left[
\begin{array}{rr}
38.75 & 1.0\\
1.0 & 38.75
\end{array}
\right]
\label{g.eq}
\end{equation}

\begin{figure*}
  \begin{center}\begin{tabular}{c c c}
    \makebox[\figwidtha][l]{\large (a)} &
    \makebox[\figwidthb][l]{\large (b)} &
    \makebox[\figwidthc][l]{\large (c)} \\
    \epsfxsize=\figwidtha \epsfbox{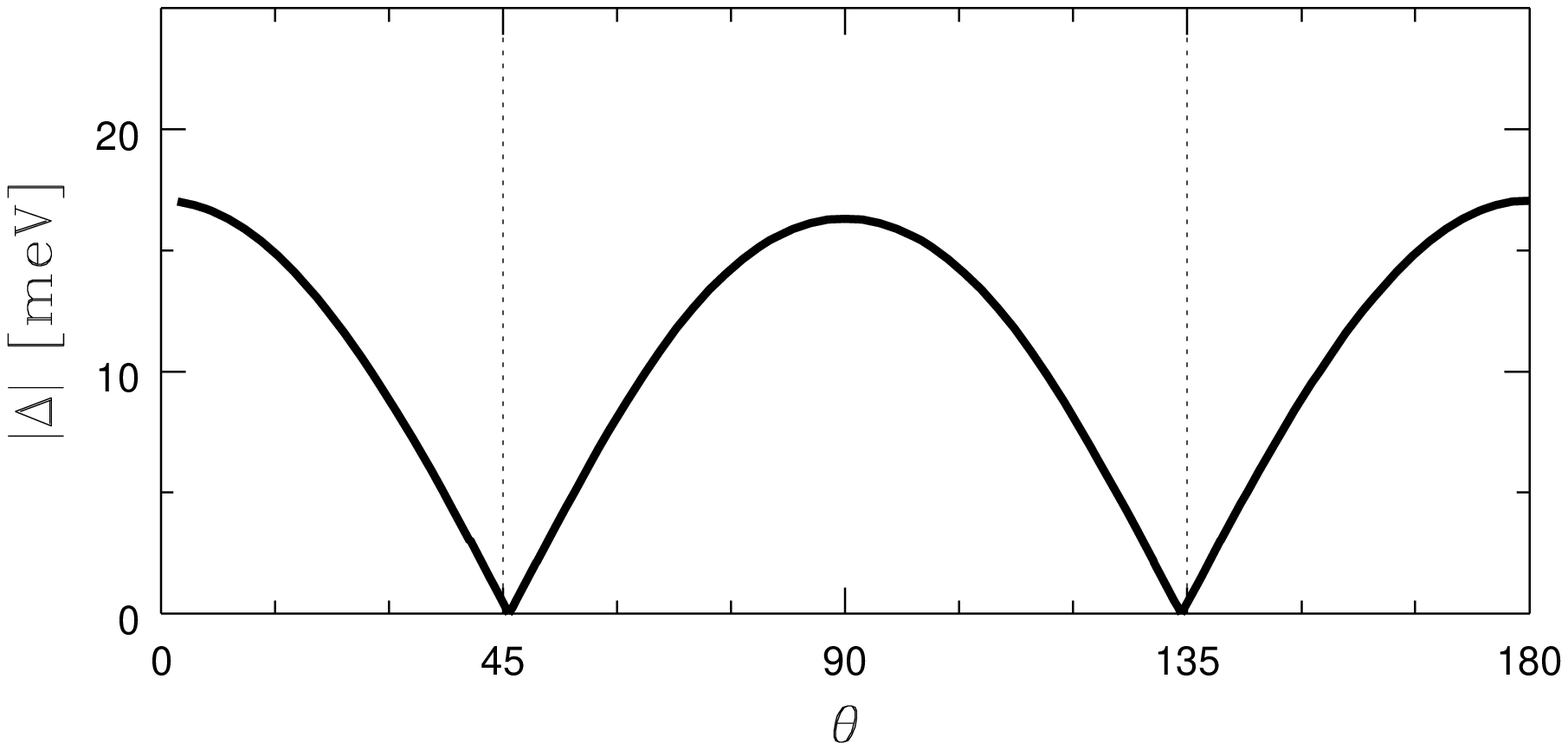} &
    \epsfxsize=\figwidthb \epsfbox{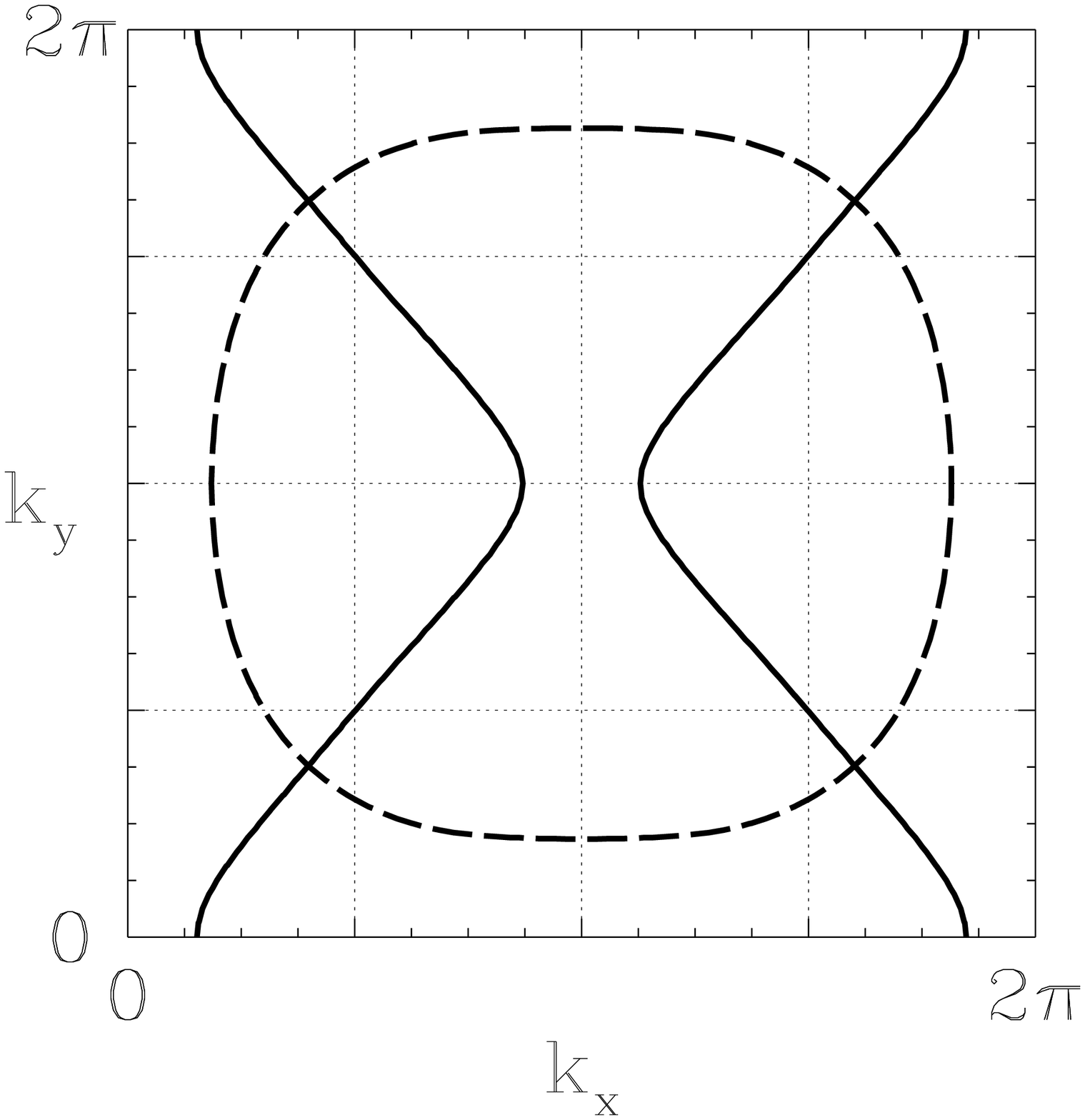} &
    \epsfxsize=\figwidthc \epsfbox{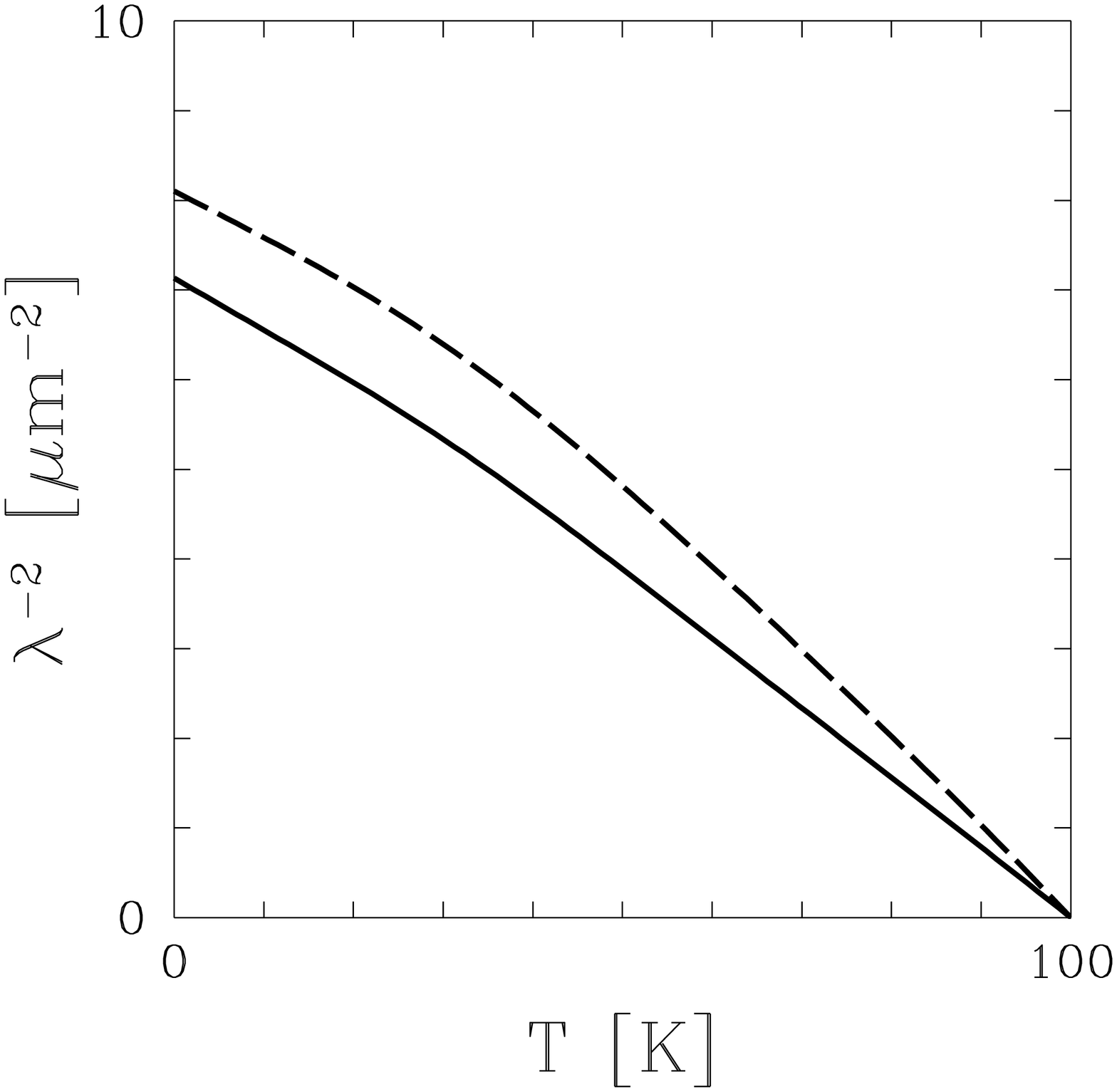} \\
    \makebox[\figwidtha][l]{\large (d)} &
    \makebox[\figwidthb][l]{\large (e)} &
    \makebox[\figwidthc][l]{\large (f)} \\
    \epsfxsize=\figwidtha \epsfbox{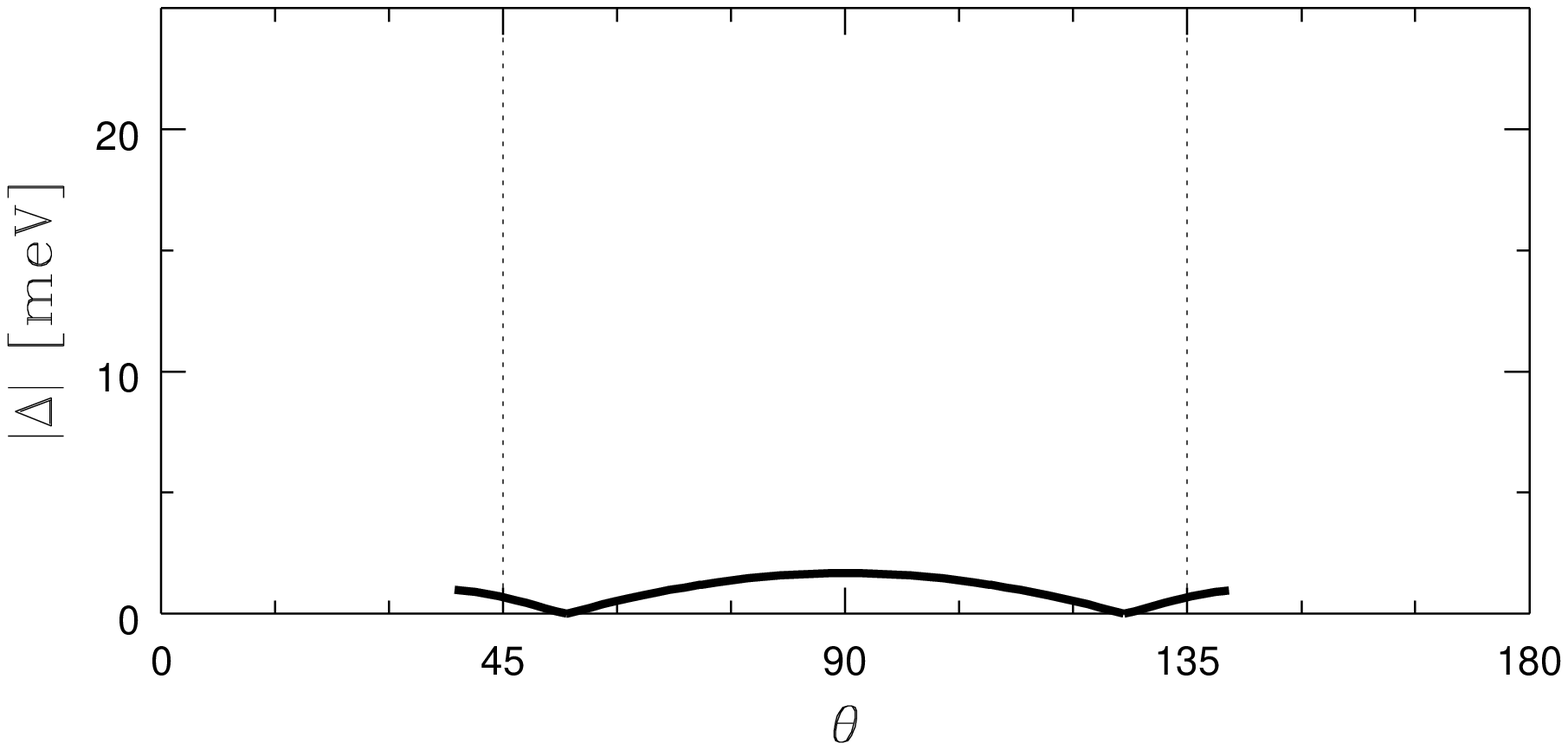} &
    \epsfxsize=\figwidthb \epsfbox{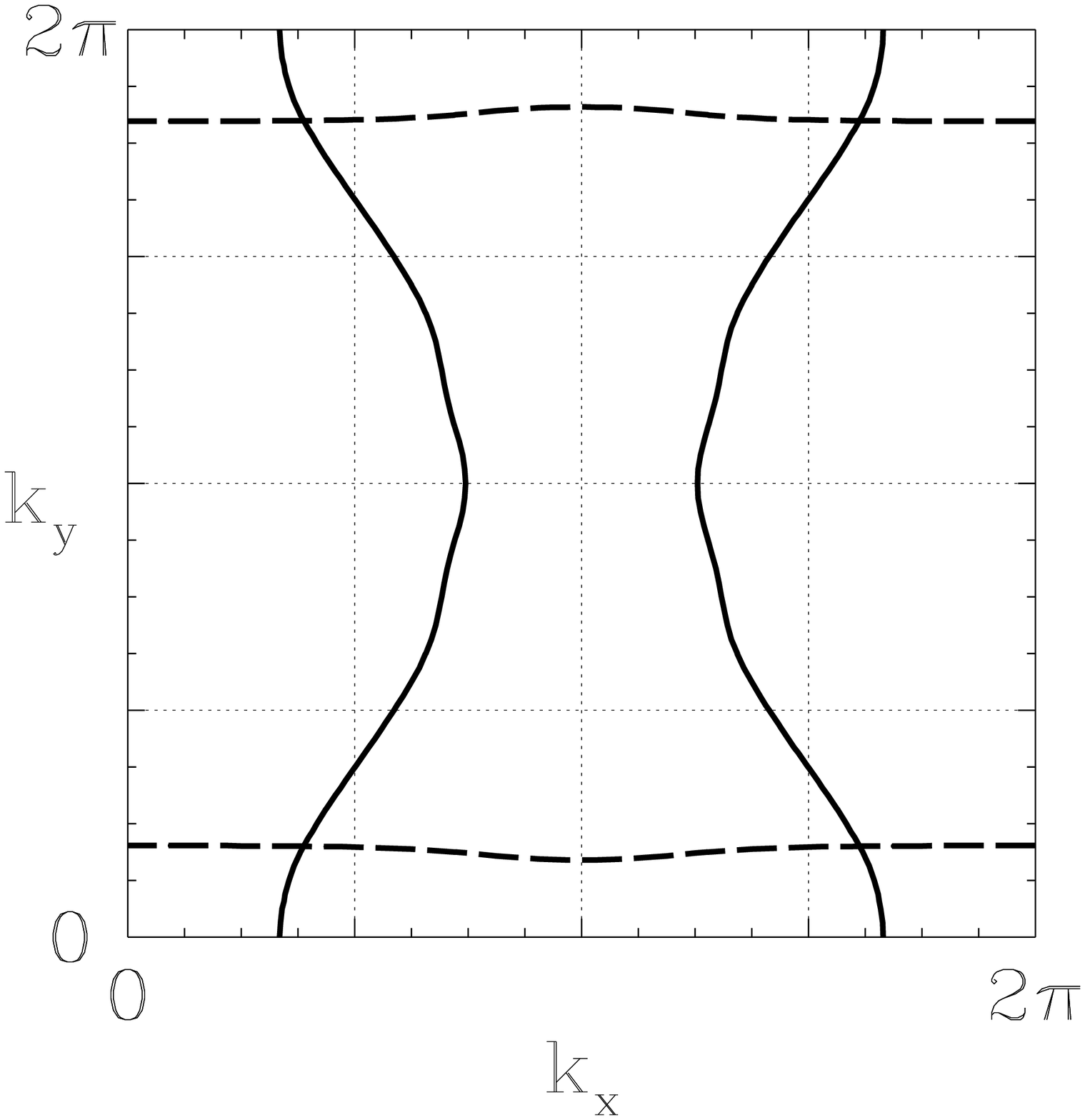} &
    \epsfxsize=\figwidthc \epsfbox{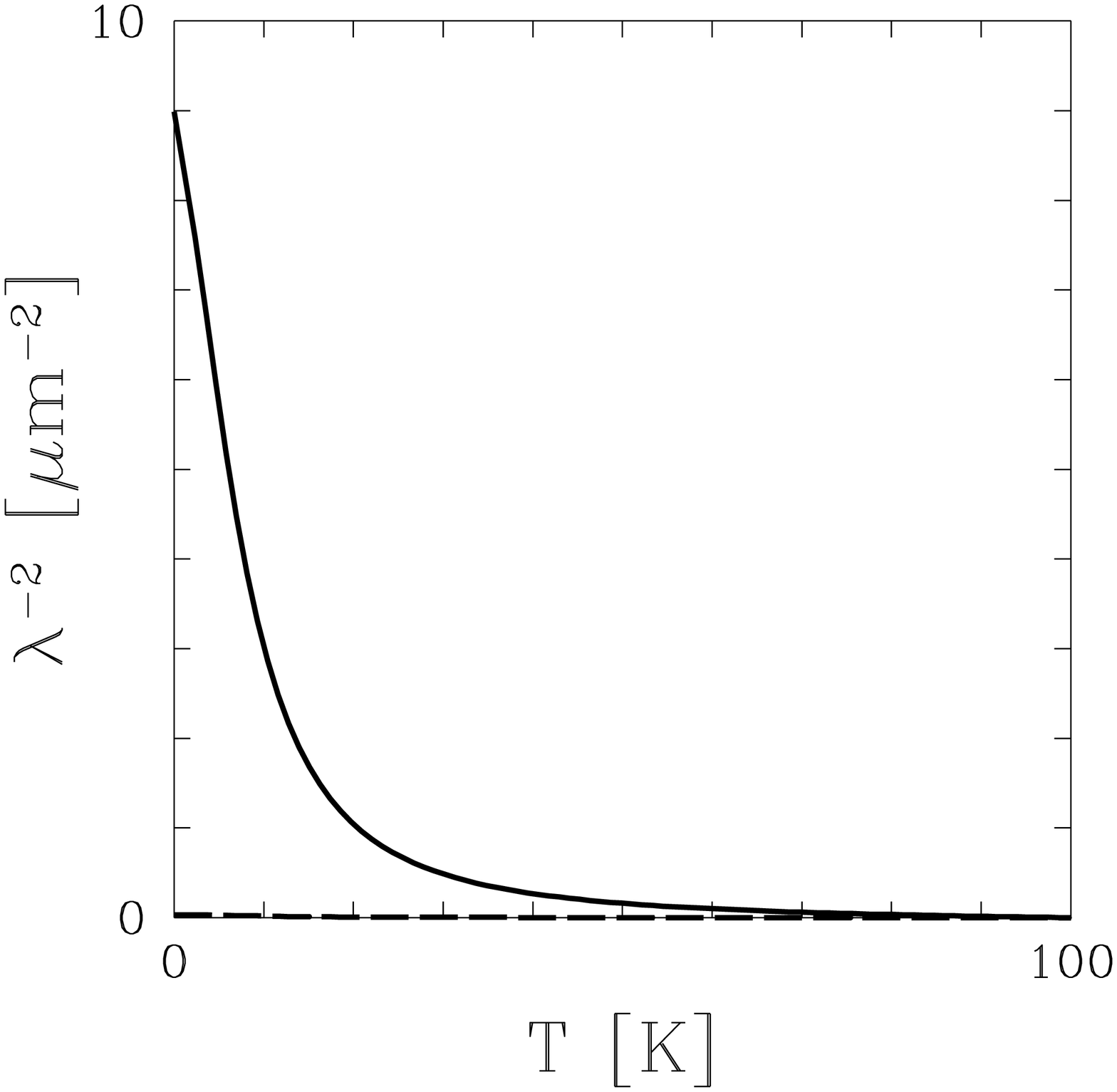}
  \end{tabular} \end{center}
\caption{
(a) and (d) are the absolute value of gap as a function of angle
$\theta$ measured clockwise from the $k_x$ axis along the Fermi
surface for the CuO$_2$ plane and CuO chains, respectively.  (b) and
(e) are Fermi surface contours (dashed line) and gap zeros (solid
line) while (c) and (f) are the contributions to the inverse square of
the penetration depth $\lambda^{-2}(T)$ as a function of temperature
along $a$- (dashed curve) and $b$-direction (solid curve).  Note that
the chains Fig. (d), (e) and (f) contribute almost exclusively to the
$b$-direction.  The same pairing was included in chains and plane
(equation \ref{g.eq}) yet note the very small energy scale for the gap
in the chains (d) as compare to the gap in the plane (a). This is due
to the quasi-one dimensional nature of the chain Fermi surface. Note
also that the low energy scale in the chains leads to the sharp drop
in penetration depth in $b$-direction with increasing temperature seen
in (f). }
\label{fig1}
\end{figure*}

Solutions of the {\sc bcs} equations using fast Fourier transforms
give an admixture of $s_\circ$, $s_{x^2+y^2}$ and $d_{x^2-y^2}$ which
all belong to the same irreducible representation for an orthorhombic
system.  The gap amplitudes are
\begin{equation}
\mit\Delta_{{\bf k}, \alpha}=
\left[\begin{array}{rrr}
-0.196 & -0.467 & 9.482\\
-0.644 & -0.182 & 1.482
\end{array}\right]
\left[\begin{array}{c}
\eta_{\bf k}^{(s_\circ)}\\
\eta_{\bf k}^{(s_x)}\\
\eta_{\bf k}^{(d)}
\end{array}\right]
\label{soln1.eq}
\end{equation}
\noindent
with the $\eta_{\bf k}^{(\cdot)}$ as given in
(\ref{so.eq}-\ref{d.eq}).  We note that in the CuO$_2$ plane, the
$d_{x^2-y^2}$ component is dominant with a small admixture of
$s_\circ$ and $s_{x^2+y^2}$ having the opposite sign (out of phase by
$\pi$).  Although the same {\sc mmp} interaction as used in the
CuO$_2$ planes has been assumed to also apply on the CuO chains, the
very different symmetry of the quasi one-dimensional Fermi surface has
led to a gap on the chains which is smaller by an order of magnitude.
The $d_{x^2-y^2}$ component has the same phase as it has in the planes
but it is now not much larger than its $s_\circ$ and $s_{x^2+y^2}$
parts.  In Fig.\ \ref{fig1}(b) (middle upper frame) and (e) (middle
lower frame), we show the CuO$_2$ and CuO Fermi surfaces (dashed
lines), respectively, as well as the contour of gap zeros (solid
lines).  In Fig.\ \ref{fig1}(b), the Fermi surface has tetragonal
symmetry but the gap zeros have nevertheless moved off the main
diagonals (solid curve) because of the admixture into the mainly
$d_{x^2-y^2}$ gap function of $s_\circ$ and $s_{x^2+y^2}$ components.
As these are negative in equation (\ref{soln1.eq}), the zeros cross
the Fermi surface at an angle ($\theta$) measured clockwise from the
negative $k_y$ axis which is slightly larger than $45^\circ$.  This is
seen more clearly in Fig.\ \ref{fig1}(a) where we show the absolute
value of the gap on the Fermi surface as a function of $\theta$ for
the plane.  Note that the gap does not display full tetragonal
symmetry in that the zeros are displaced from 45$^\circ$ and
135$^\circ$ and that the gap at 90$^\circ$ is smaller than it is at
0$^\circ$ and 180$^\circ$.

\begin{figure}
  \postscript{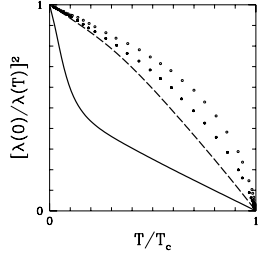}
\caption{
Total contribution to the $a$- (dashed) and $b$-directions (solid) as
a function of temperature to the inverse square of the penetration
depth normalized to its zero temperature value.  The dots are
experimental values\protect\cite{basov} with $b$-direction falling below the
$a$-direction.  The $b$-direction theoretical curve is much too steep
at low temperature because of the small energy scale for the chain
gap. }
\label{fig2}
\end{figure}

     For the chain case, the gap zeros (solid curves in Fig.\
\ref{fig1}(e)) are displaced much more off the main diagonals and as
illustrated in Fig.\ \ref{fig1}(d) the value of the absolute value of
the gap as a function of $\theta$ along the quasi one-dimensional
chain Fermi surface is very small.  Note that this is a result of the
orthorhombicity (ie, geometry of the Fermi surface) of the chain Fermi
surface as the pairing potential has been assumed to have the same
size and form in the chains as in the tetragonal planes.  For the
given value of pairing interaction, the gap ratio $2\mit\Delta_{\rm
max}/T_{\rm c}$ is 3.86 in the plane while it has been reduced to 0.47
in the chains.  Finally, it is important to note that in Fig.\
\ref{fig1}(d), the gap on the Fermi surface starts at a finite value
of $\theta$ reflecting the geometry of the nearly linear chain Fermi
surface.  The very small energy scale exhibited by the gap in the
chains leads directly to a very steep slope in the resulting $b$-axis
penetration depth as seen in the solid curve of Fig.\ \ref{fig1}(f).
The contribution of the chains to the $a$-axis penetration depth is by
contrast (dashed line) very small.  This is quite different from the
contribution of the plane to the penetration depth illustrated in
Fig.\ \ref{fig1}(c).  In this case, the nearly tetragonal symmetry of
the Fermi surface leads to very similar contributions for the
temperature dependence of the $a$- (dashed) and $b$- (solid) direction
penetration depth.  The small differences between these two curves as
to absolute value of $\lambda^{-2}(0)$ at zero temperature is due to
the orthorhombic symmetry of the dispersion (equation
(\ref{disp.eq})).  Unfortunately, these small differences are not
measured directly in an experiment which probes the sum of
contributions from chains and planes, ie, Fig.\ \ref{fig1}(f) and
\ref{fig1}(c).  This sum is shown in Fig.\ \ref{fig2} where it is compared with
experiments\cite{basov} (circles).  The agreement is poor although the
order of the zero temperature slopes is correct, i.e. it is steeper in
the chain direction ($b$-direction).  The large discrepancy can be
traced directly to the small energy scale found in the chains in this
model and we can conclude that the data favours a gap which is of the
same magnitude in the chains as it is in the planes.  Finally, we note
that $\lambda_a/\lambda_b\simeq 1.4$ in our calculation, which is
reasonably close to (but not exactly) the measured value.

\begin{figure*}
  \begin{center}\begin{tabular}{c c c}
    \makebox[\figwidtha][l]{\large (a)} &
    \makebox[\figwidthb][l]{\large (b)} &
    \makebox[\figwidthc][l]{\large (c)} \\
    \epsfxsize=\figwidtha \epsfbox{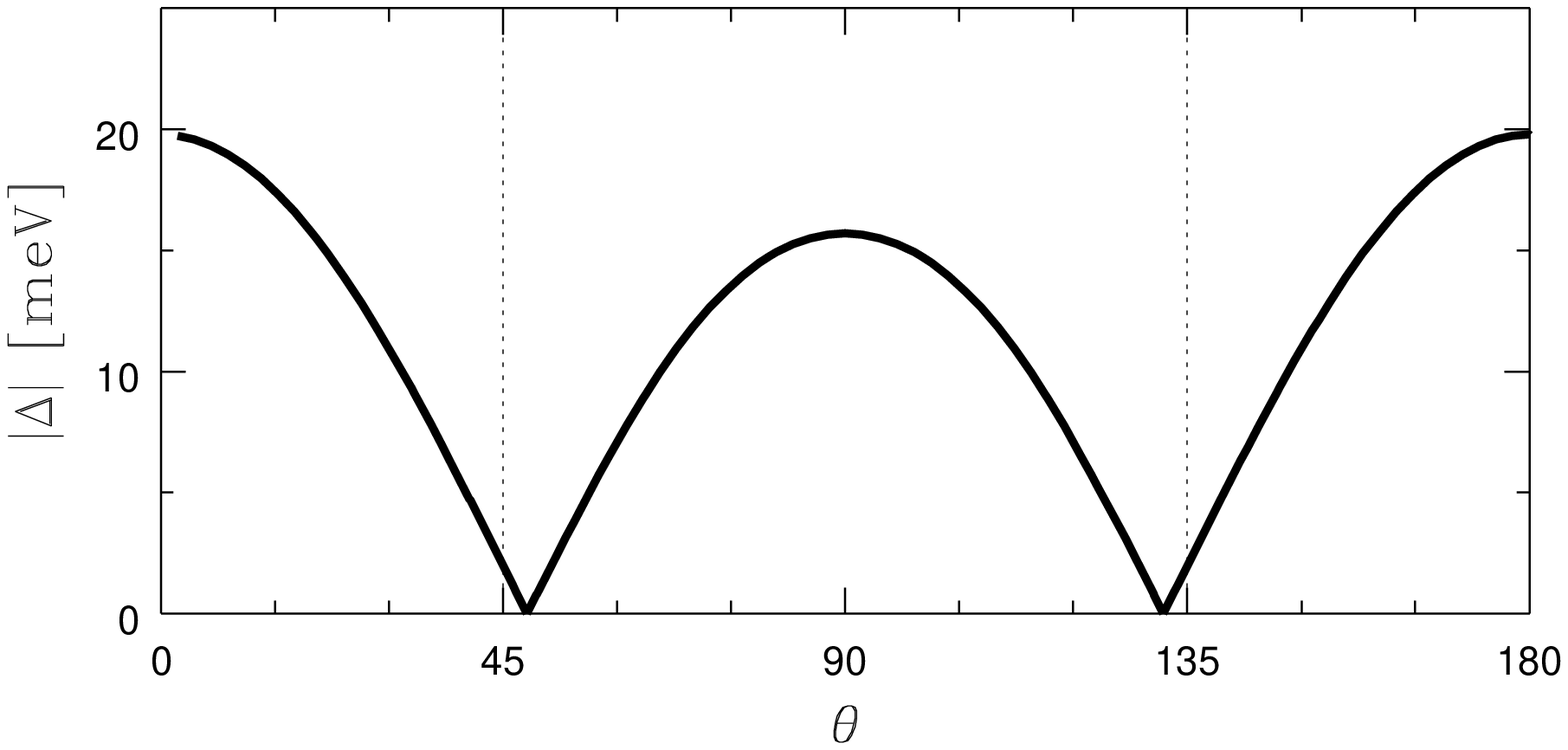} &
    \epsfxsize=\figwidthb \epsfbox{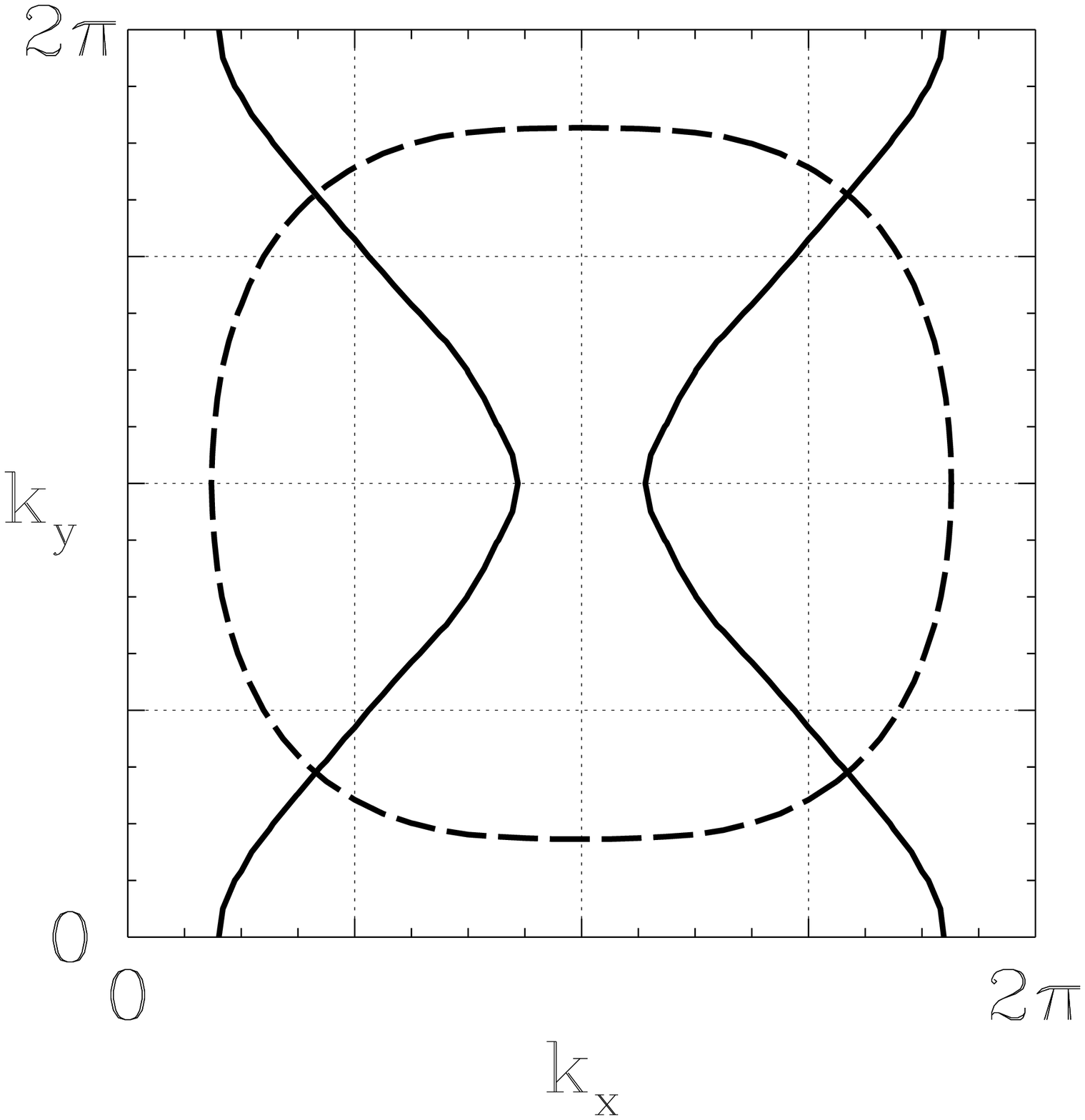} &
    \epsfxsize=\figwidthc \epsfbox{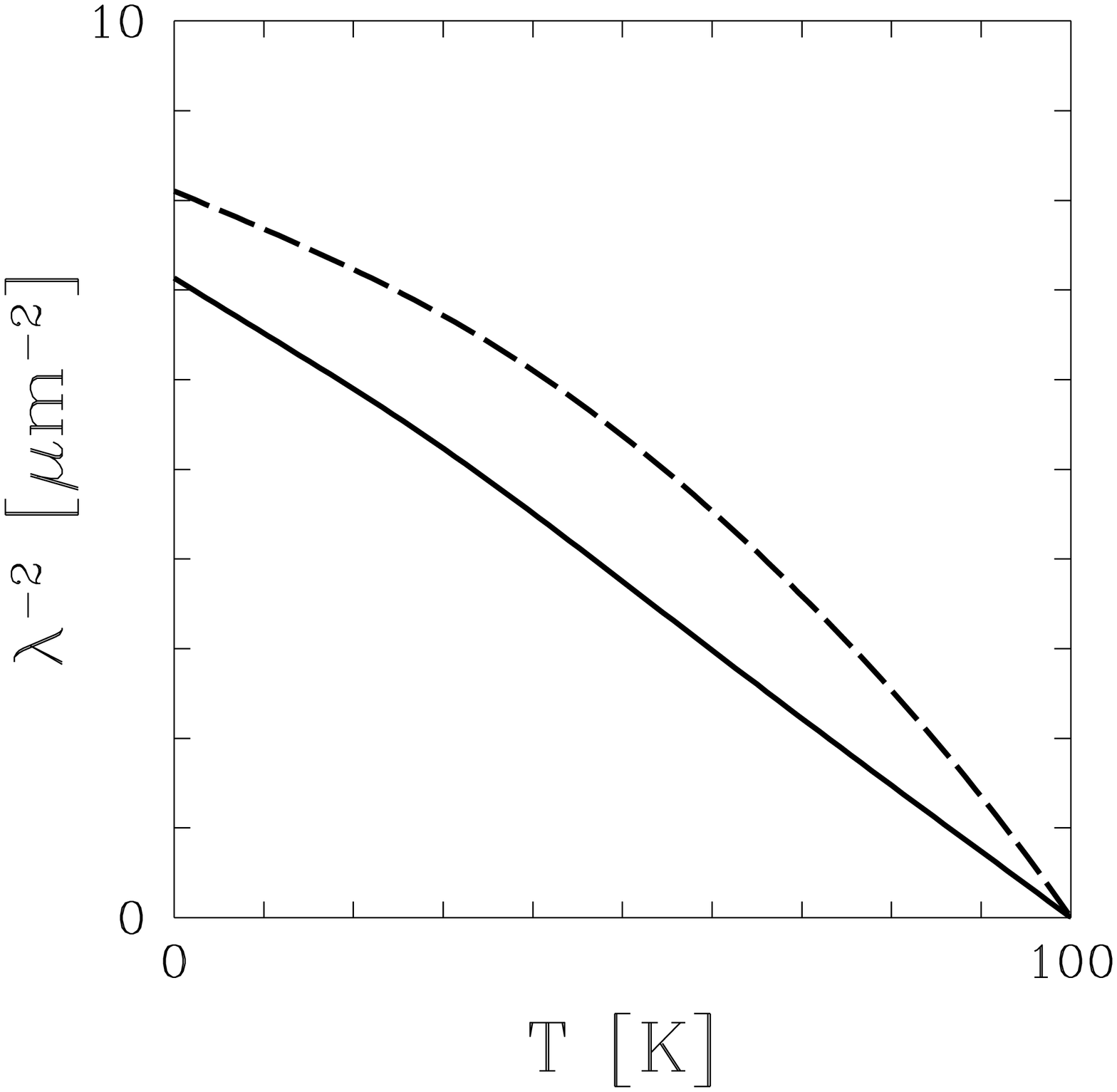} \\
    \makebox[\figwidtha][l]{\large (d)} &
    \makebox[\figwidthb][l]{\large (e)} &
    \makebox[\figwidthc][l]{\large (f)} \\
    \epsfxsize=\figwidtha \epsfbox{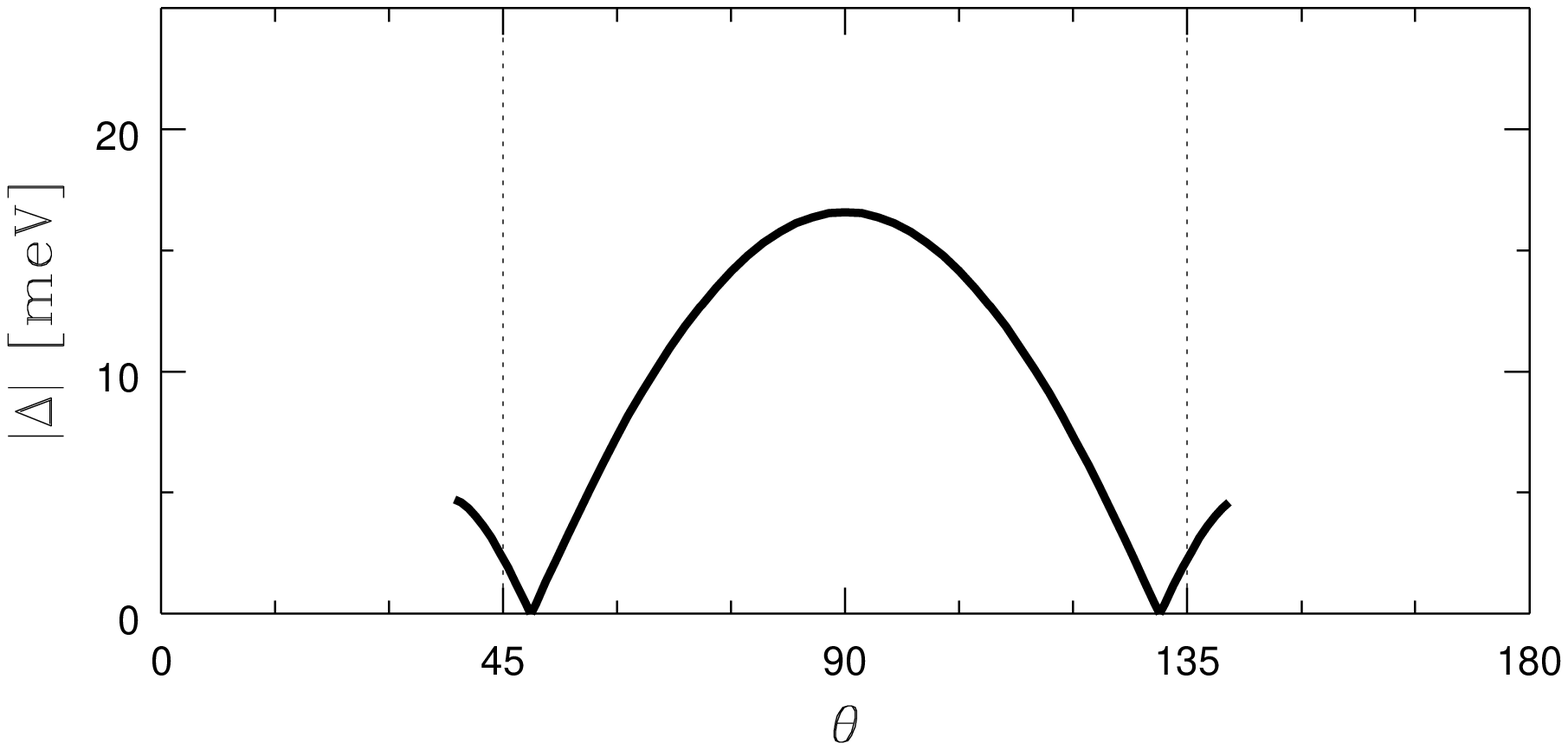} &
    \epsfxsize=\figwidthb \epsfbox{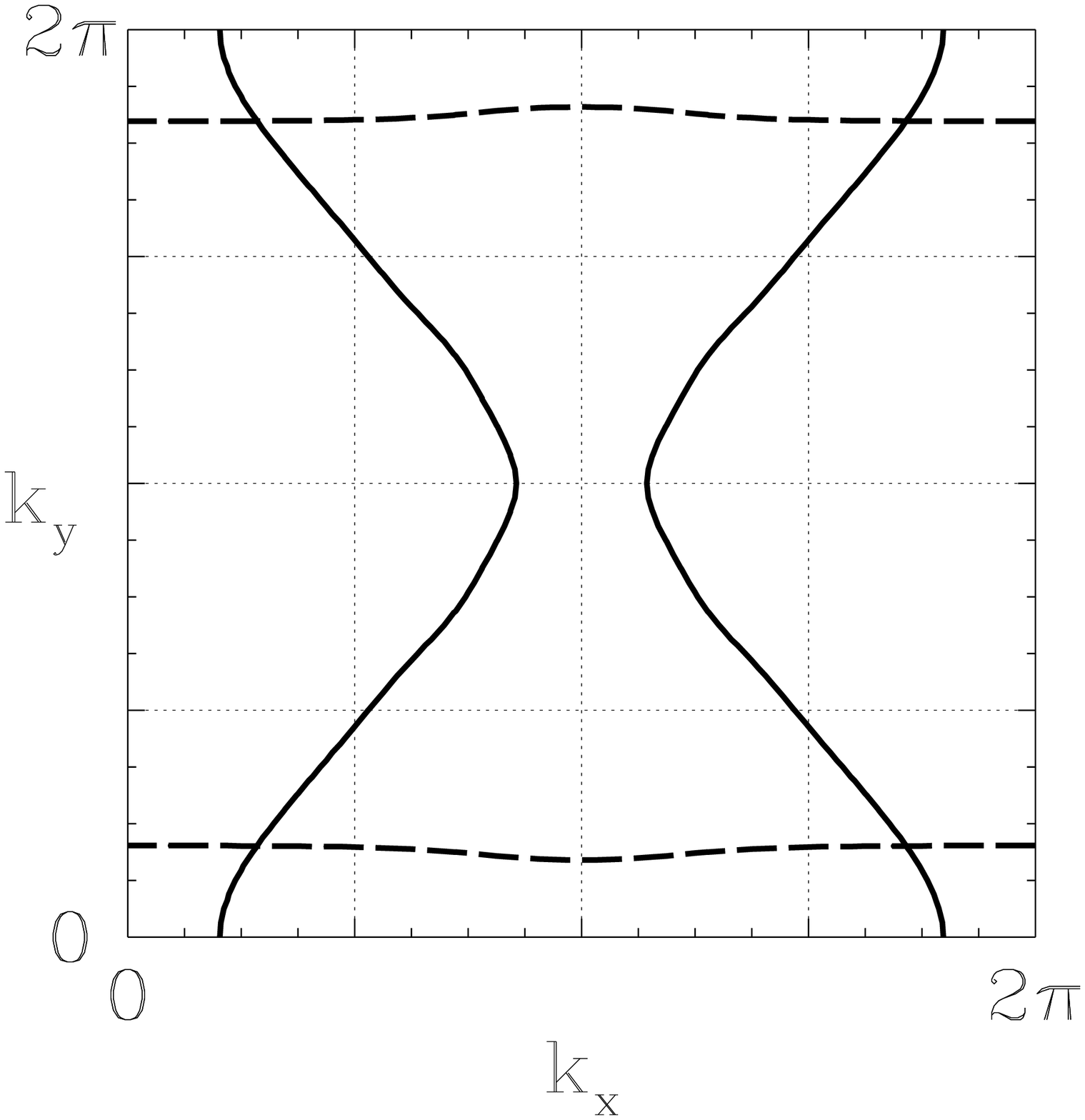} &
    \epsfxsize=\figwidthc \epsfbox{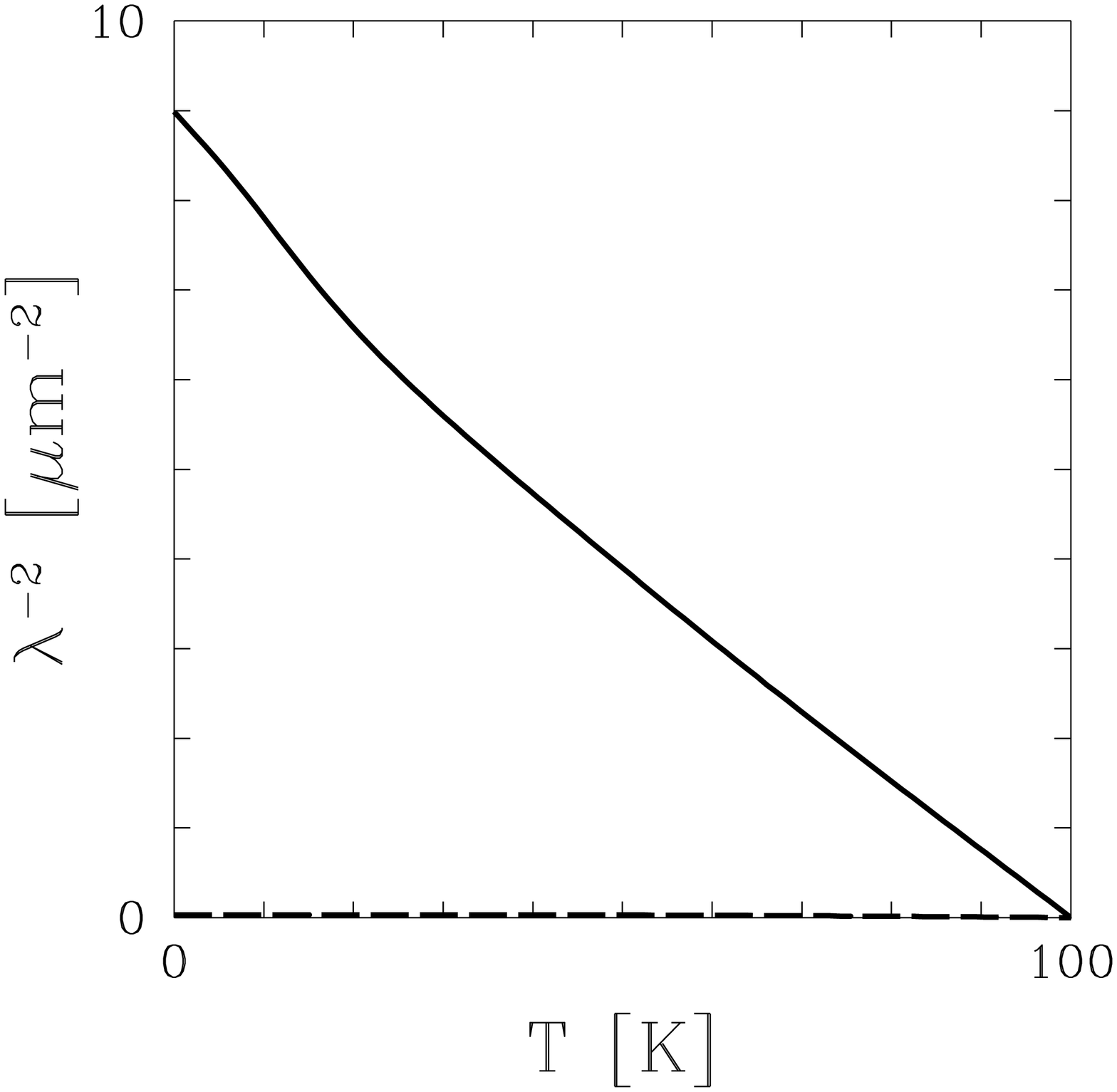}
  \end{tabular} \end{center}
\caption{
Frame (a), (b) and (c) apply to the CuO$_2$ planes while (d), (e) and
(f) apply to the CuO chains with a nearly one dimensional Fermi
surface.  The pairing interaction is assumed to be the same on chains
and planes and off diagonal so that a single gap results.  Figs. (b)
and (e) give the Fermi surface contours (dashed curve) and the gap
zeros (solid curves) which are the same in both frames but different
from pure $d$-wave which would be lines on the main diagonals in the
Brillouin zone.  Frame (a) and (d) give the absolute value of the gap
as a function of angle $\theta$ along the Fermi surface measured
clockwise from the $k_x$ axis.  Frame (c) and (f) give partial
contributions to the inverse square of the penetration depth as a
function of temperature along $a$-direction (dashed) and $b$-direction
(solid).  Note that the zeros in the gap are not at 45$^\circ$ and
135$^\circ$ as they would be in pure $d$-wave (Fig. (a) and (d)). }
\label{fig3}
\end{figure*}

     It is easy to modify our model so that chains and planes have the
same gap value.  A single gap results for $\mit\Delta_{{\bf
k},\alpha}$ independent of $\alpha$ if all diagonal and off diagonal
elements of the pairing potential are taken to have the same value,
i.e. $g_{\alpha\beta}=$ constant for any $\alpha$ and $\beta$.  In
Fig.\ \ref{fig3}, we illustrate results obtained with the same band
parameters as used in our previous model but now with
$g_{\alpha\beta}=24.7$ for all $\alpha$ and $\beta$ in which case a
single gap results which has the form:\begin{equation}
\mit\Delta_{\bf k}=
\left[\begin{array}{rrr}
-1.83 & -0.288 & 10.09
\end{array}\right]
\left[\begin{array}{c}
\eta_{\bf k}^{(s_\circ)}\\
\eta_{\bf k}^{(s_x)}\\
\eta_{\bf k}^{(d)}
\end{array}\right].
\label{soln2.eq}
\end{equation}

This gap has mainly $d$-wave character with a small admixture of
$s_\circ$ and $s_{x^2+y^2}$ so that the gap zeros (solid curve) in
Fig.\ \ref{fig3}(b) and \ref{fig1}(e) have moved off the main diagonal
as in the previous example.  Now these zero contours are the same in
both chains and planes.  The absolute value of the gap (on the Fermi
surface) as a function of $\theta$ are shown in Fig.\ \ref{fig3}(a)
and (d) for plane and chain, respectively, and the maximum value of
$2\mit\Delta/k_{\rm B}T_{\rm c}$ is 4.5 in the planes and 3.76 in the
chains, much more comparable in value than was the case in the first
model considered.  It is to be noted that over the entire Brillouin
zone the gaps in the planes and chains are not different but, of
course, the Fermi surface geometry is totally different in these two
cases.  In particular, we note that on the Fermi surface the gap in
the chains starts at a finite angle $\theta$ near 45$^\circ$ and that
there is no Fermi surface at small values of $\theta$.

\begin{figure}
  \postscript{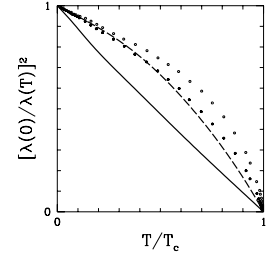}
\caption{
Total contribution to the $a$- (dashed) and $b$-directions (solid) as
a function of temperature to the inverse square of the penetration
depth normalized to its zero temperature value.  The dots are
experimental values\protect\cite{basov} with the $b$-direction falling
below the $a$-direction.  The agreement is good and value of
$\lambda_a/\lambda_b\approx 1.4$ is also in reasonable but not exact
agreement with experiment. }
\label{fig4}
\end{figure}

     The smaller value of the maximum gap found in the chains as
opposed to the planes will lead to a steeper slope of the low
temperature dependence of the inverse square of the penetration depth
in the chain direction as seen in Fig.\ \ref{fig4} where we compare with
experiments.  While theory predicts somewhat steeper slopes,
qualitative agreement with experiment is obtained in this simple
model.

\begin{figure}
  \postscript{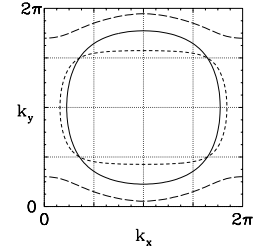}
\caption{
The three Fermi surfaces in the triplanar model of YBCO.  The chain
like band (dashed curve) is nearly one dimensional, the odd band Fermi
surface has orthorhombic symmetry (dotted curve) due to its
interaction with the chains and the even band (solid curve) remains
tetragonal. }
\label{fig5}
\end{figure}

     We turn next to results obtained for the trilayer model with
CuO$_2$ planes on both sides of a CuO chain plane.  The transverse
coupling between CuO$_2$ and CuO plane is denoted by $t_1$ and $t_2$
is a direct transverse coupling between the two CuO$_2$ planes of the
same unit cell.  For $\{t_\circ, t_1, t_2\}=\{50, 30, 15\}$ in meV the
resulting Fermi surfaces obtained from equation \ref{hamiltonian.eq}
are shown in Fig.\ \ref{fig5} where:
\begin{equation}
\left\{\begin{array}{rrrr}
t_\circ & \epsilon_\alpha & B_\alpha & \mu_\alpha
\end{array}\right\}
=
\left\{\begin{array}{cccc}
50.0 & 0.0 & 0.45 & 0.51\\
50.0 & 1.0 & 0.0 & 1.2\\
50.0 & 0.0 & 0.45 & 0.51
\end{array}\right\},
\label{params2.eq}
\end{equation}
\noindent
with $t_\circ$ in meV.  The chain Fermi surface, denoted by the
dashed line, is quasi one dimensional and has only orthorhombic
symmetry.  The two CuO$_2$ plane Fermi surfaces are shown as dotted
(odd) and solid (even) curves.  Only the even band Fermi surface has
tetragonal symmetry while the odd band is flatter along $k_x$ than it
is along $k_y$.  This orthorhombicity will lead to a different $a$-
and $b$-direction contributions to the penetration depth from the
CuO$_2$ planes themselves and this will now be described.  

\begin{figure*}
  \begin{center}\begin{tabular}{c c c}
    \makebox[\figwidtha][l]{\large (a)} &
    \makebox[\figwidthb][l]{\large (b)} &
    \makebox[\figwidthc][l]{\large (c)} \\
    \epsfxsize=\figwidtha \epsfbox{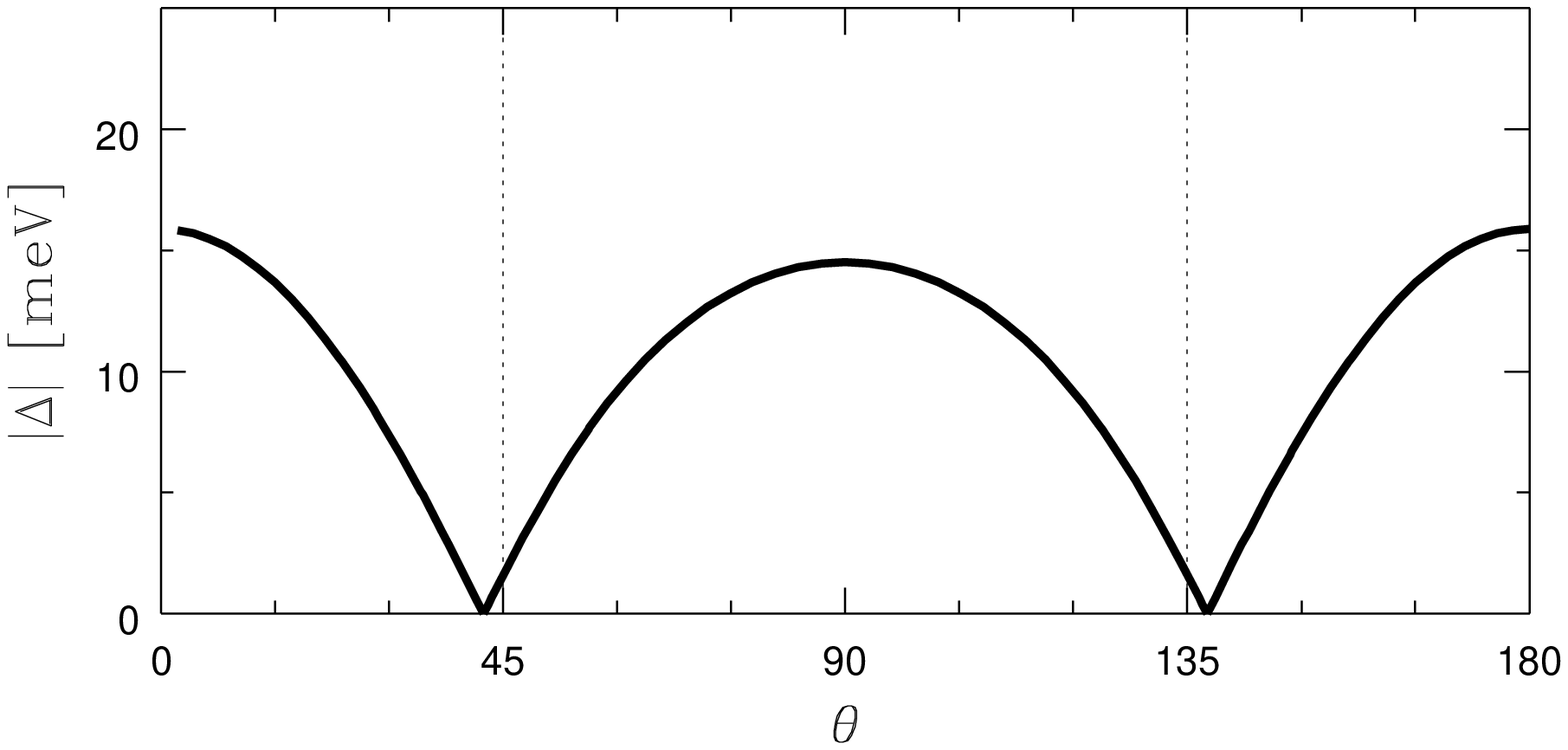} &
    \epsfxsize=\figwidthb \epsfbox{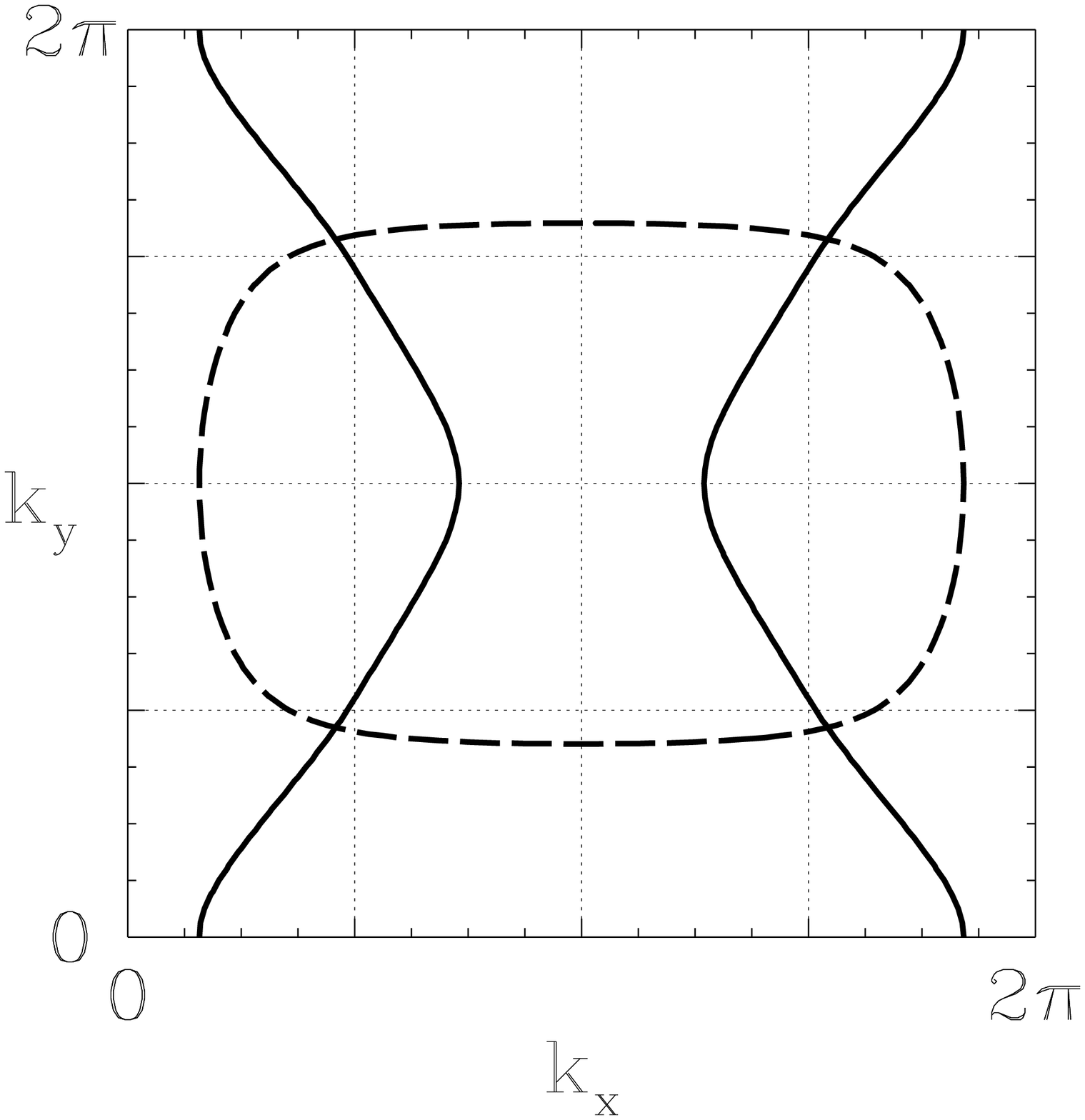} &
    \epsfxsize=\figwidthc \epsfbox{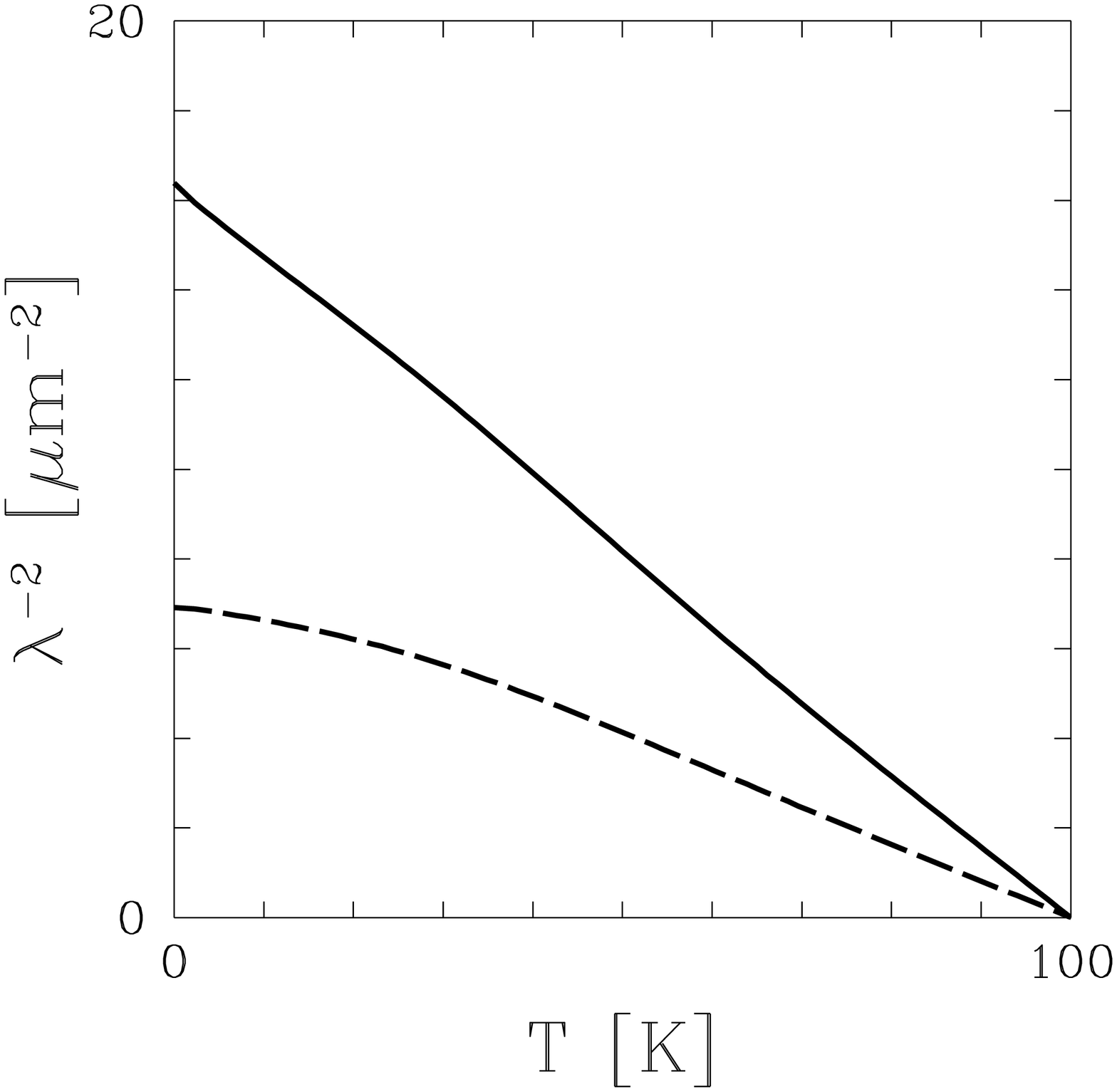} \\
    \makebox[\figwidtha][l]{\large (d)} &
    \makebox[\figwidthb][l]{\large (e)} &
    \makebox[\figwidthc][l]{\large (f)} \\
    \epsfxsize=\figwidtha \epsfbox{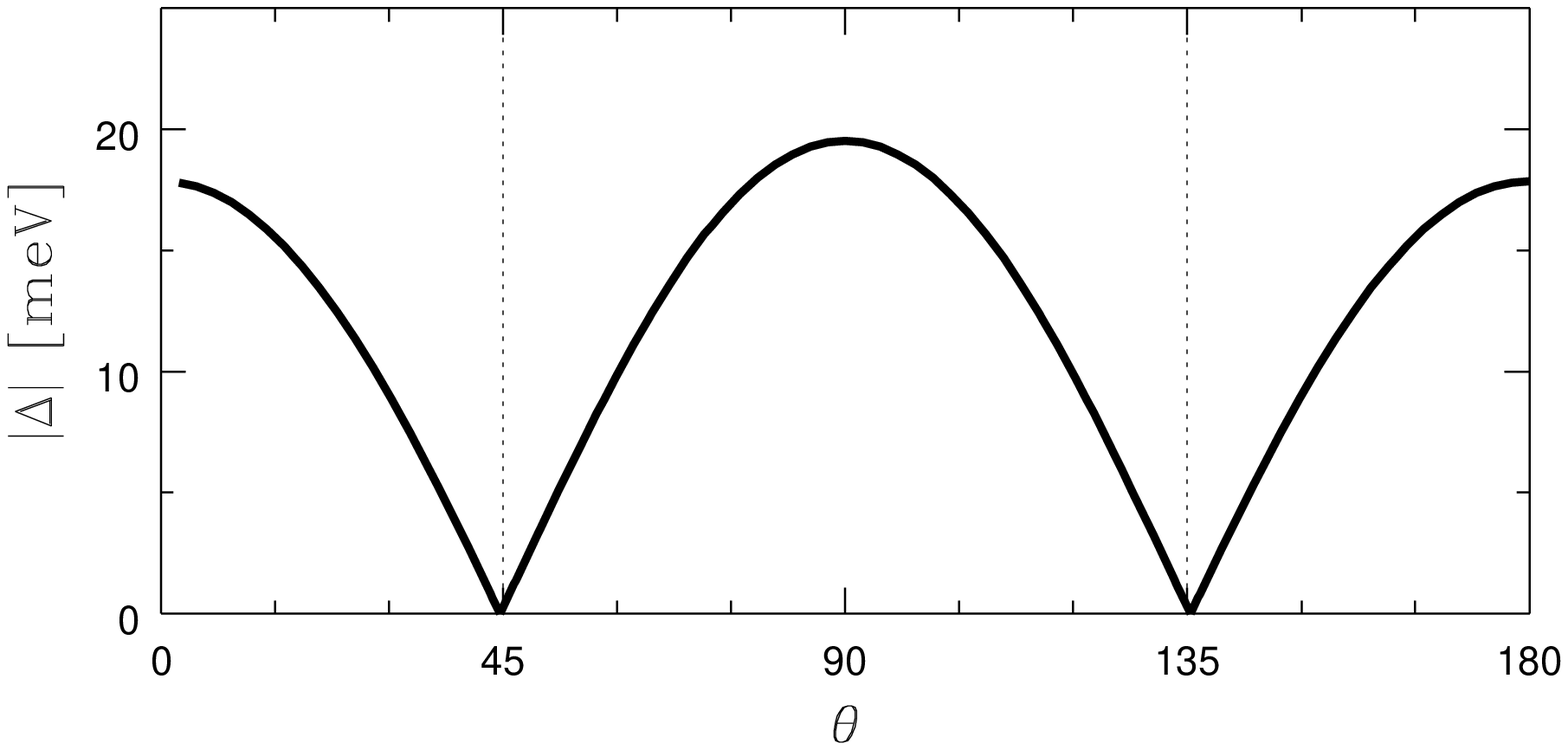} &
    \epsfxsize=\figwidthb \epsfbox{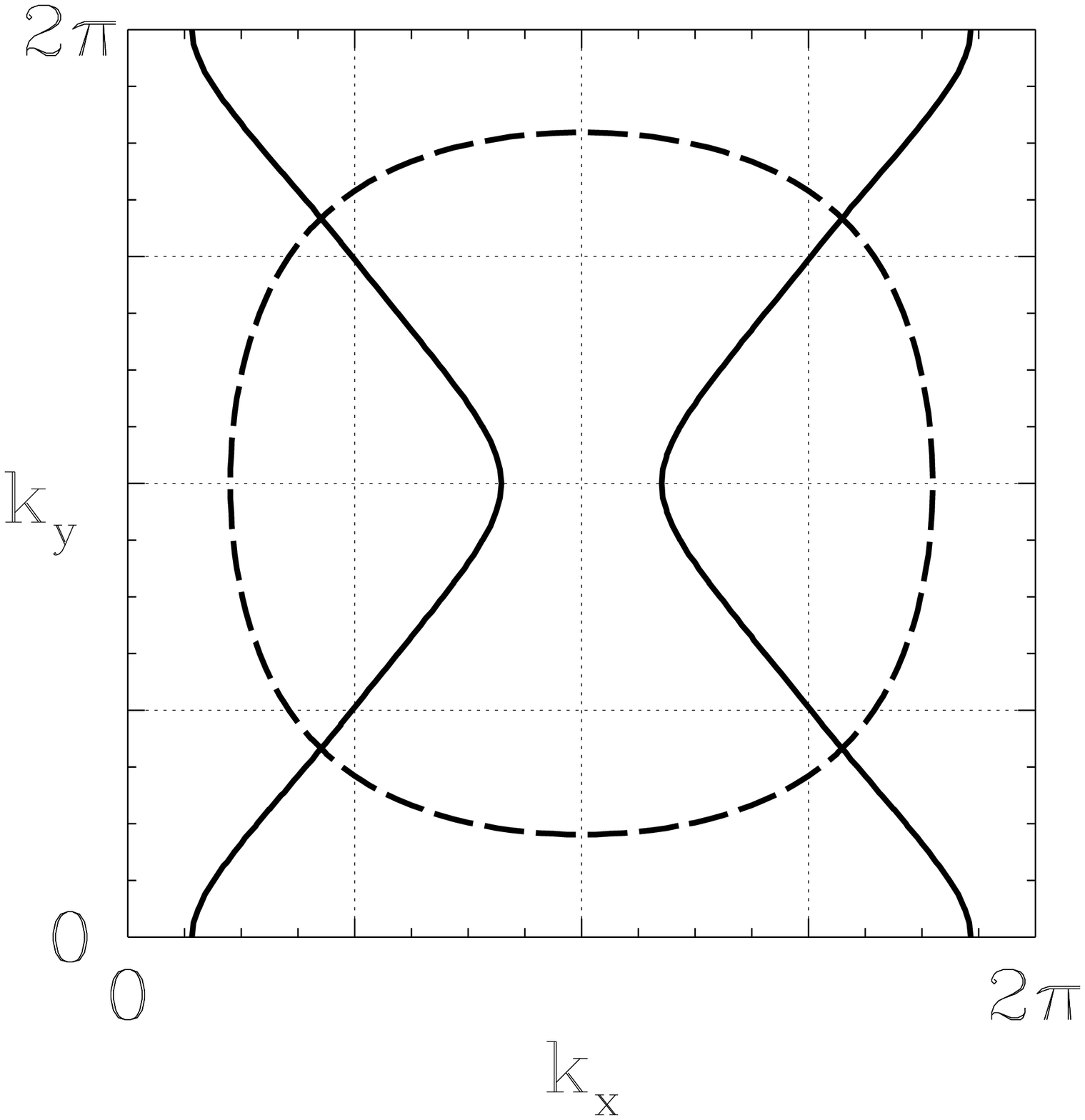} &
    \epsfxsize=\figwidthc \epsfbox{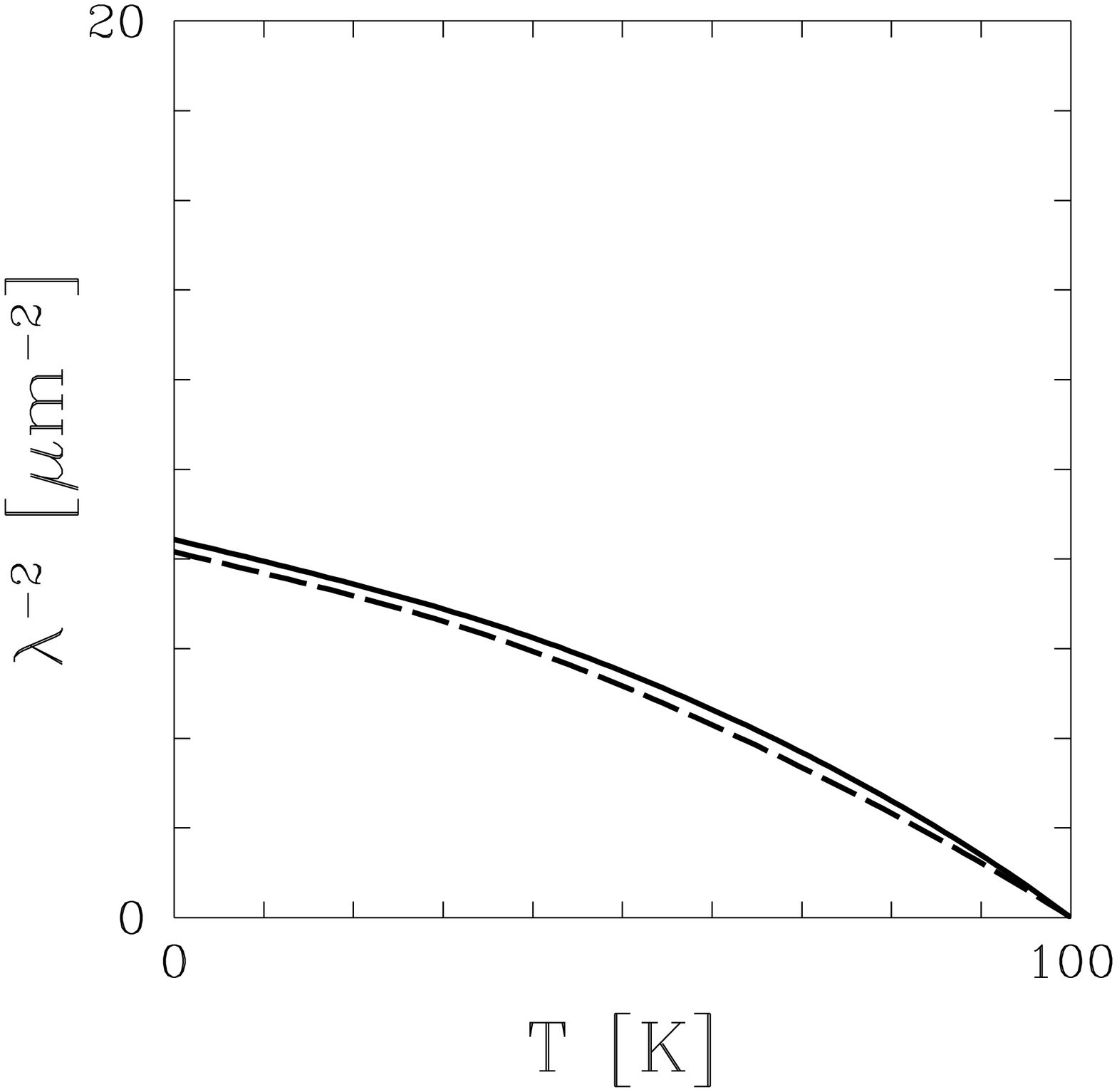}
  \end{tabular} \end{center}
\caption{
Results for the triplanar model illustrated in Fig.\
\protect\ref{fig5} assuming no pairing diagonal or off diagonal on the
chains which then remain normal.  Frame (a), (b) and (c) apply to the
odd band (orthorhombic symmetry) while (d), (e) and (f) apply to the
even band (tetragonal symmetry).  The first two frames (a) and (d)
illustrate the absolute value of the gap along the Fermi surface as a
function of angle $\theta$ measured clockwise from the $k_x$ axis.
These curves do not have full tetragonal symmetry and the zeros are
displaced somewhat from the ideal $d_{x^2-y^2}$ case which would be at
45$^\circ$ and 135$^\circ$.  Frame (b) and (e) give the Fermi surface
(dashed line) and gap nodes (solid line) for odd and even plane
(tetragonal), respectively.  The final frames (c) and (f) give partial
contributions to the inverse square of the penetration depth as a
function of temperature for $a$- (dashed) and $b$- (solid) direction.
Note that the odd (orthorhombic) band gives a contribution to the
$b$-direction which is about twice its contribution to the
$a$-direction, while the even band (tetragonal) gives nearly the same
contribution in both directions.  In the model, the ratio
$\lambda_a/\lambda_b\approx 1.3$. }
\label{fig6}
\end{figure*}

A first set of results that is instructive is illustrated in Fig.\ \ref{fig6}.
The chains are assumed not to contribute directly to the penetration
depth but only provide an agency whereby the CuO$_2$ plane acquires
orthorhombic symmetry.  This is accomplished easily if no pairing is
assumed to act on the chains be it diagonal or off diagonal.  That is,
we take for the pairing potential the matrix:
\begin{equation}
g_{\alpha\beta}=
\left[\begin{array}{ccc}
36.8 & 0.0 & 15.0\\
0.0 & 0.0 & 0.0\\
15.0 & 0.0 & 36.8
\end{array}\right].
\label{interaction1.eq}
\end{equation}

What is critical in equation (\ref{interaction1.eq}) is that the
second row is all zero so the chains remain normal by arrangement.  In
this case, we obtain a gap of the form:
\begin{equation}
\mit\Delta_{{\bf k},\alpha}=
\left[\begin{array}{ccc}
1.81 & -1.46 & 9.96\\
0.0 & 0.0 & 0.0\\
0.59 & -0.80 & 10.88
\end{array}\right]
\left[\begin{array}{c}
\eta_{\bf k}^{(s_\circ)}\\
\eta_{\bf k}^{(s_x)}\\
\eta_{\bf k}^{(d)}
\end{array}\right].
\label{soln3.eq}
\end{equation}

\begin{figure}
  \postscript{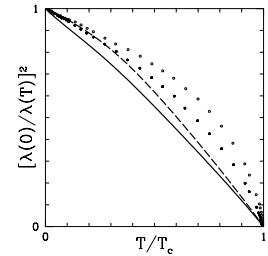}
\caption{
Total contributions to the inverse square of the penetration depth
normalized to its zero temperature value as a function of temperature.
The solid curve is for the $b$-direction and the dashed for the
$a$-direction.  The dots are the experimental data\protect\cite{basov}
with $b$-direction curve below the $a$-direction curve.  Results are
for the trilayer model of Figs.\ \protect\ref{fig5} and \ref{fig6} in which it
is assumed that there is no pairing on the chains on or off diagonal
so they remain normal and do not contribute directly to the
condensate. }
\label{fig7}
\end{figure}

The gap on even (top row) and odd (bottom row) band is mainly $d$-wave
like with $s$-components an order of magnitude smaller.  The contours
of gap zeros are illustrated as the solid curves in Fig.\
\ref{fig6}(b) and (e) for odd and even bands, respectively.  The Fermi
contour is the dashed curve in each case.  The absolute value of the
gap ($|\mit\Delta|$) as a function of angle $\theta$ along the Fermi
surface are given in Fig.\ \ref{fig5}(a) and (d) for odd and even
band, respectively.  In both cases, the gap zeros occur slightly
before $\theta=45^\circ$ and beyond 135$^\circ$ so that the gap does
not have tetragonal symmetry even in the even band with a tetragonal
Fermi surface.  Also, the maximum gap on the Fermi surface is not
quite the same in the two bands and is $2\mit\Delta/k_{\rm B}T_{\rm
c}=$4.5 and 3.69 for even and odd cases, respectively.  While these
values are not very different, the contributions of even and odd bands
to the $a$- and $b$-direction penetration depths are very different.
The odd band with orthorhombic Fermi surface gives a factor of 2
difference in the value of the zero temperature penetration depth as
shown in Fig.\ \ref{fig6}(c).  The solid curve applies along the
$b$-direction and the dashed curve along the $a$-direction.  For the
even Fermi surface, which has tetragonal symmetry, the situation is
very different as illustrated in Fig.\ \ref{fig6}(f) where it is seen
that solid (along $b$-direction) and dashed curve (along
$a$-direction) are nearly the same (the slight difference is due to
the difference in the $a$ and $b$ lattice parameters).  Finally, in
Fig.\ \ref{fig7}, we compare results for the $a$- (dashed) and
$b$-direction (solid) penetration depth with experimental results.
Again the curves are correctly ordered as to value of slope at zero
temperature and the agreement is qualitative.  It is as good as that
shown in Fig.\ \ref{fig4}, although the physics of the two cases is
quite different.  In Fig.\ \ref{fig7}, the chains are normal by
assumption and do not contribute directly to the condensate density in
the $a$- or $b$-directions.  The in plane anisotropy is due mainly to
the odd band Fermi surface which has been made orthorhombic through
its coupling to the chains and much less importantly to the $s$
admixture in the gap function.  Thus, we are dealing mainly with a
band structure effect which dominates and overwhelms any gap symmetry
effects.

\begin{figure*}
  \begin{center}\begin{tabular}{c c c}
    \makebox[\figwidtha][l]{\large (a)} &
    \makebox[\figwidthb][l]{\large (b)} &
    \makebox[\figwidthc][l]{\large (c)} \\
    \epsfxsize=\figwidtha \epsfbox{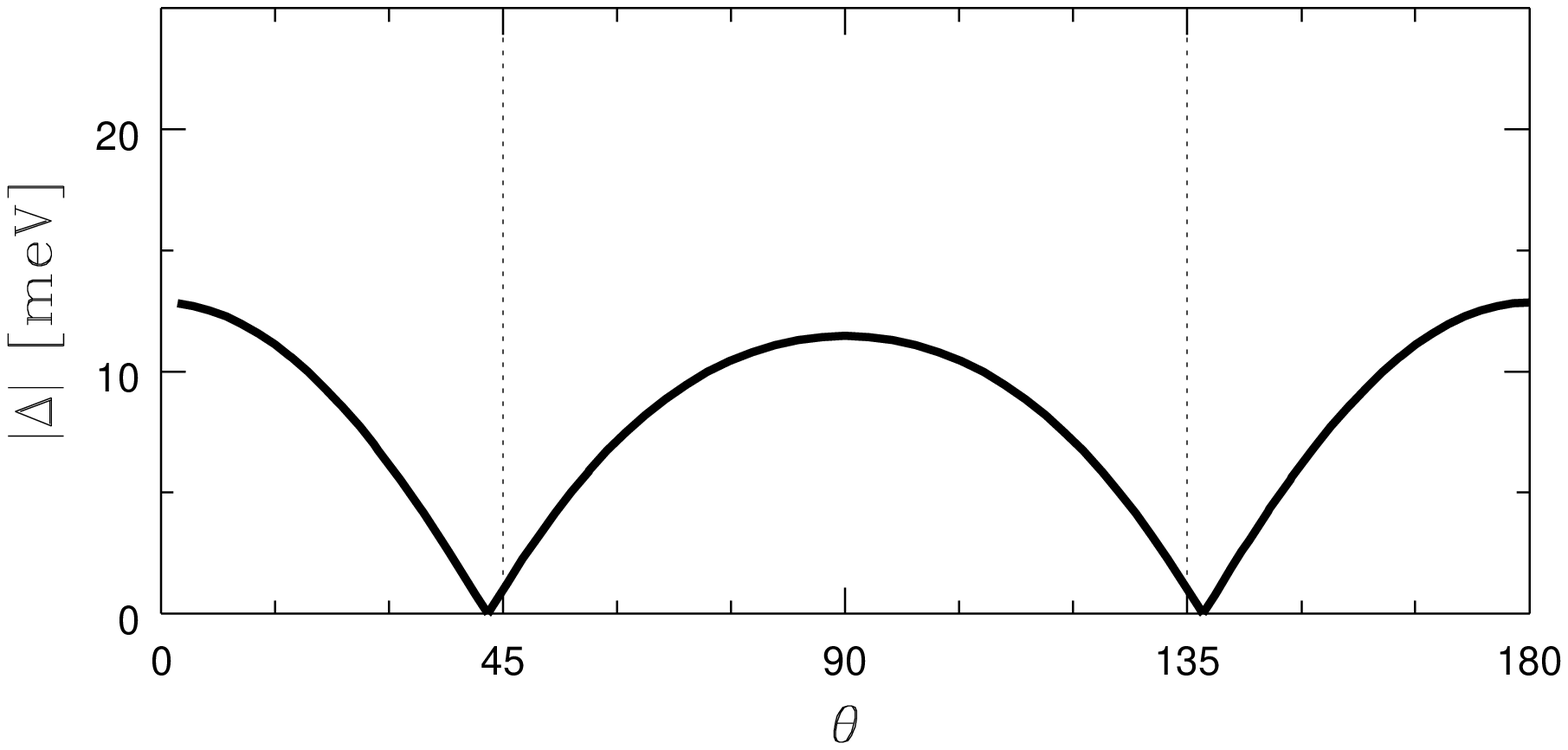} &
    \epsfxsize=\figwidthb \epsfbox{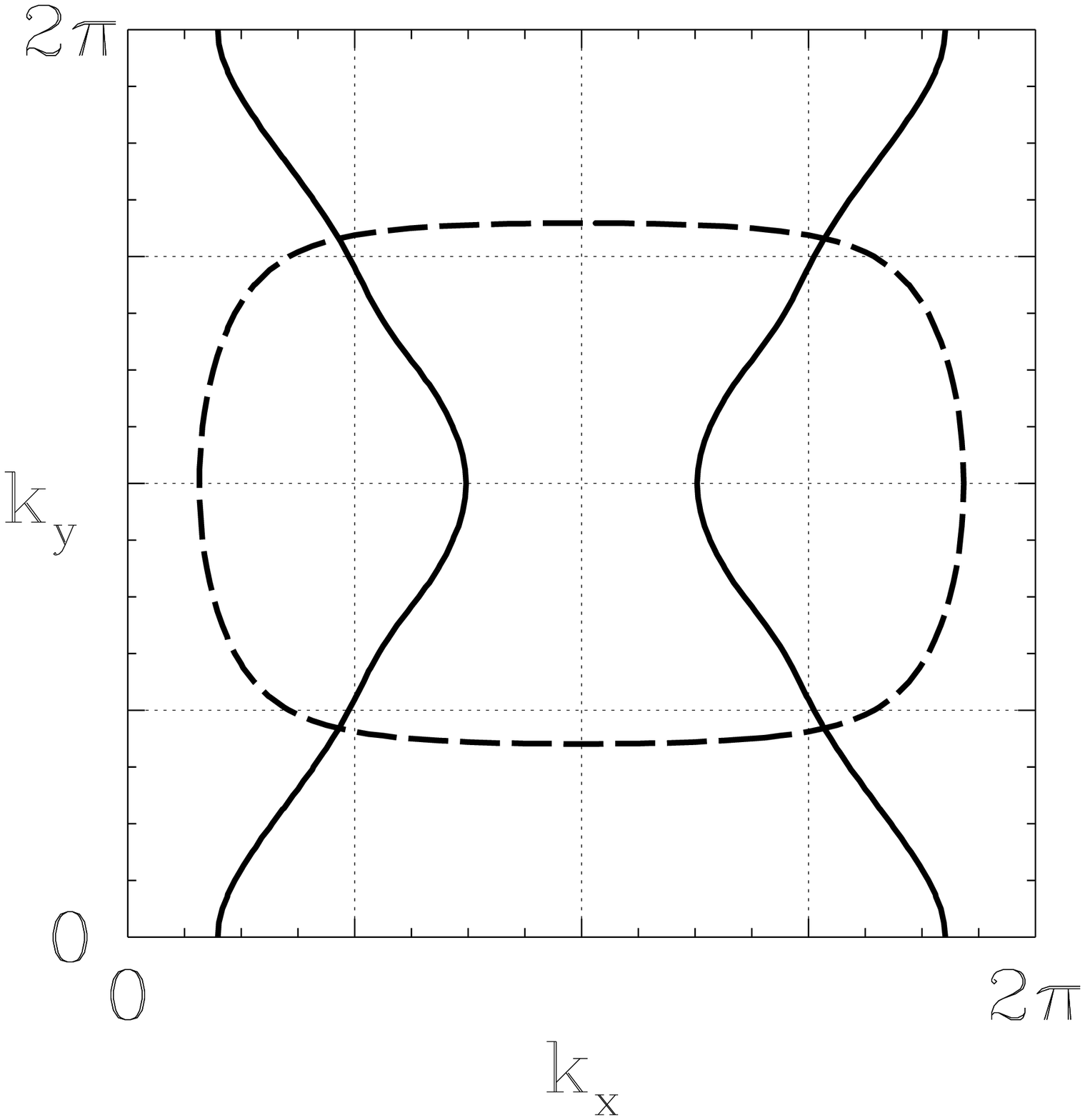} &
    \epsfxsize=\figwidthc \epsfbox{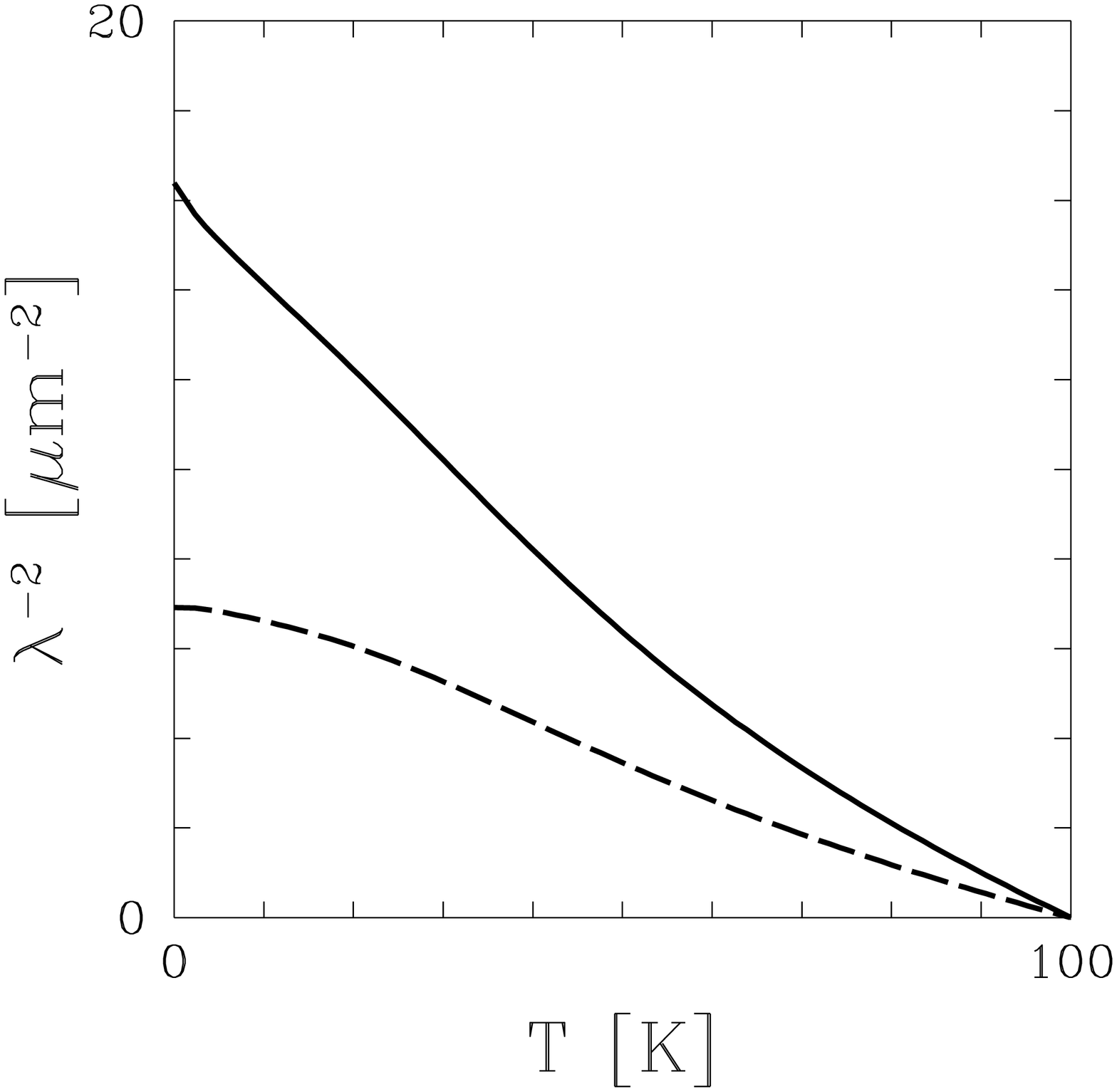} \\
    \makebox[\figwidtha][l]{\large (d)} &
    \makebox[\figwidthb][l]{\large (e)} &
    \makebox[\figwidthc][l]{\large (f)} \\
    \epsfxsize=\figwidtha \epsfbox{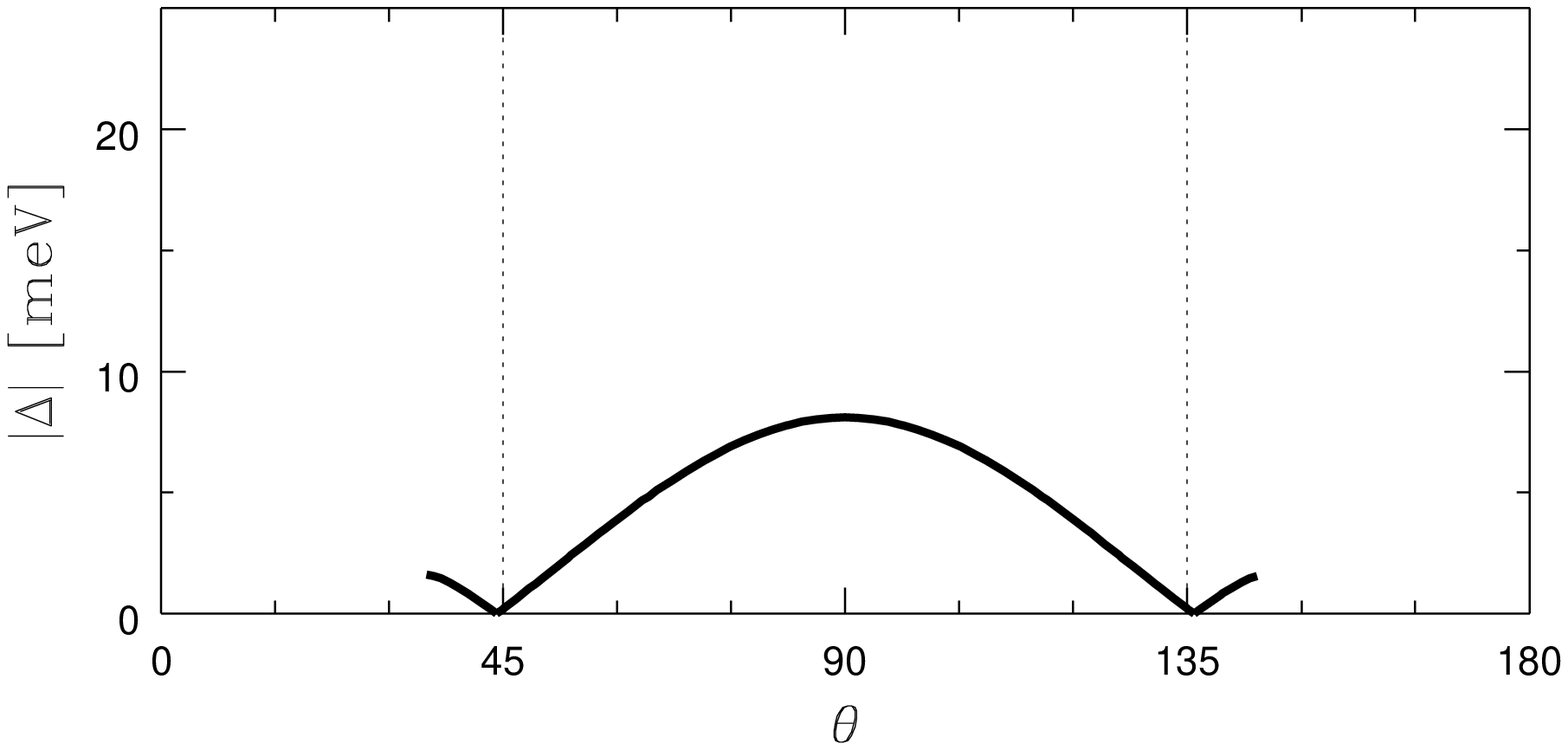} &
    \epsfxsize=\figwidthb \epsfbox{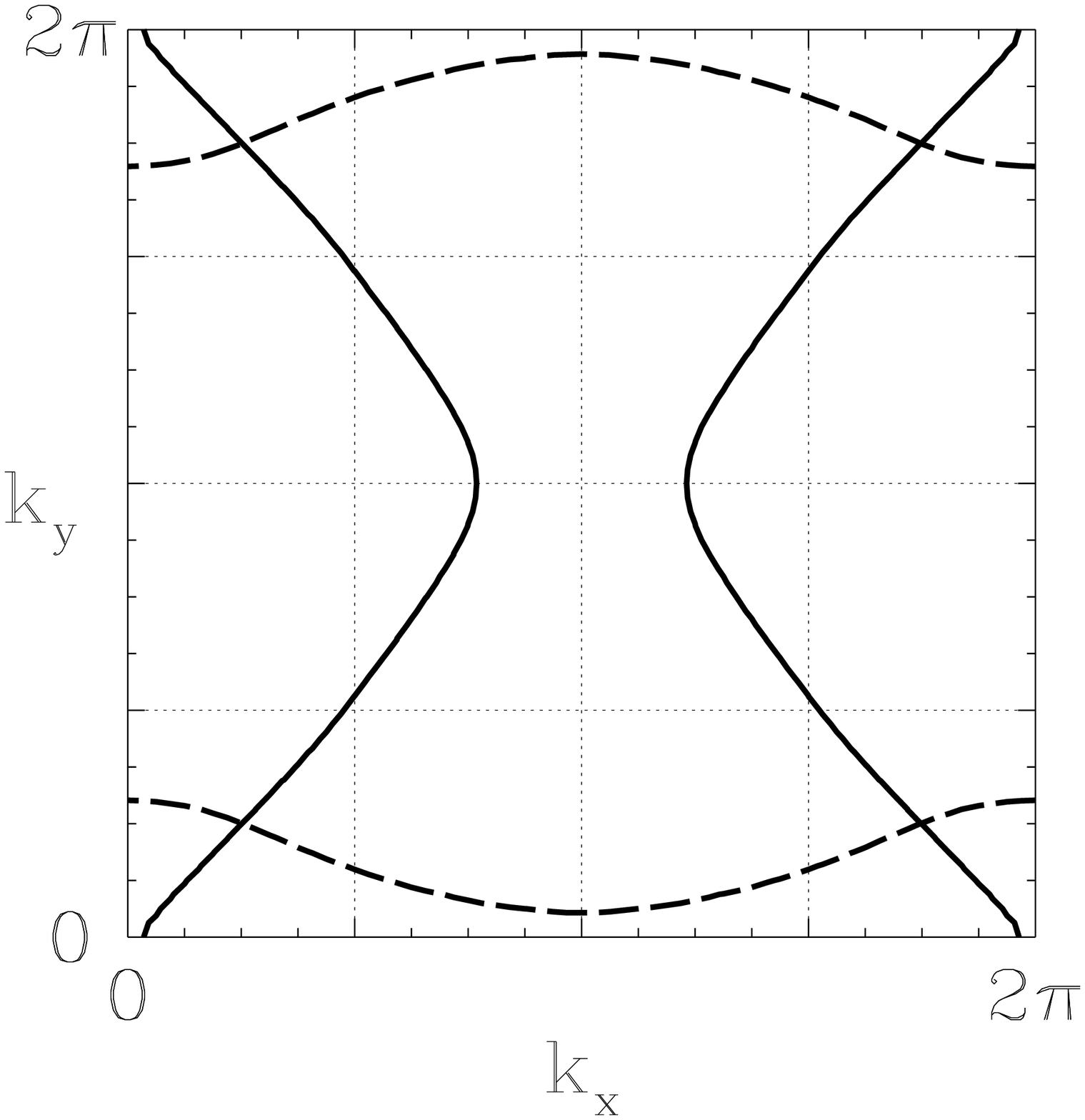} &
    \epsfxsize=\figwidthc \epsfbox{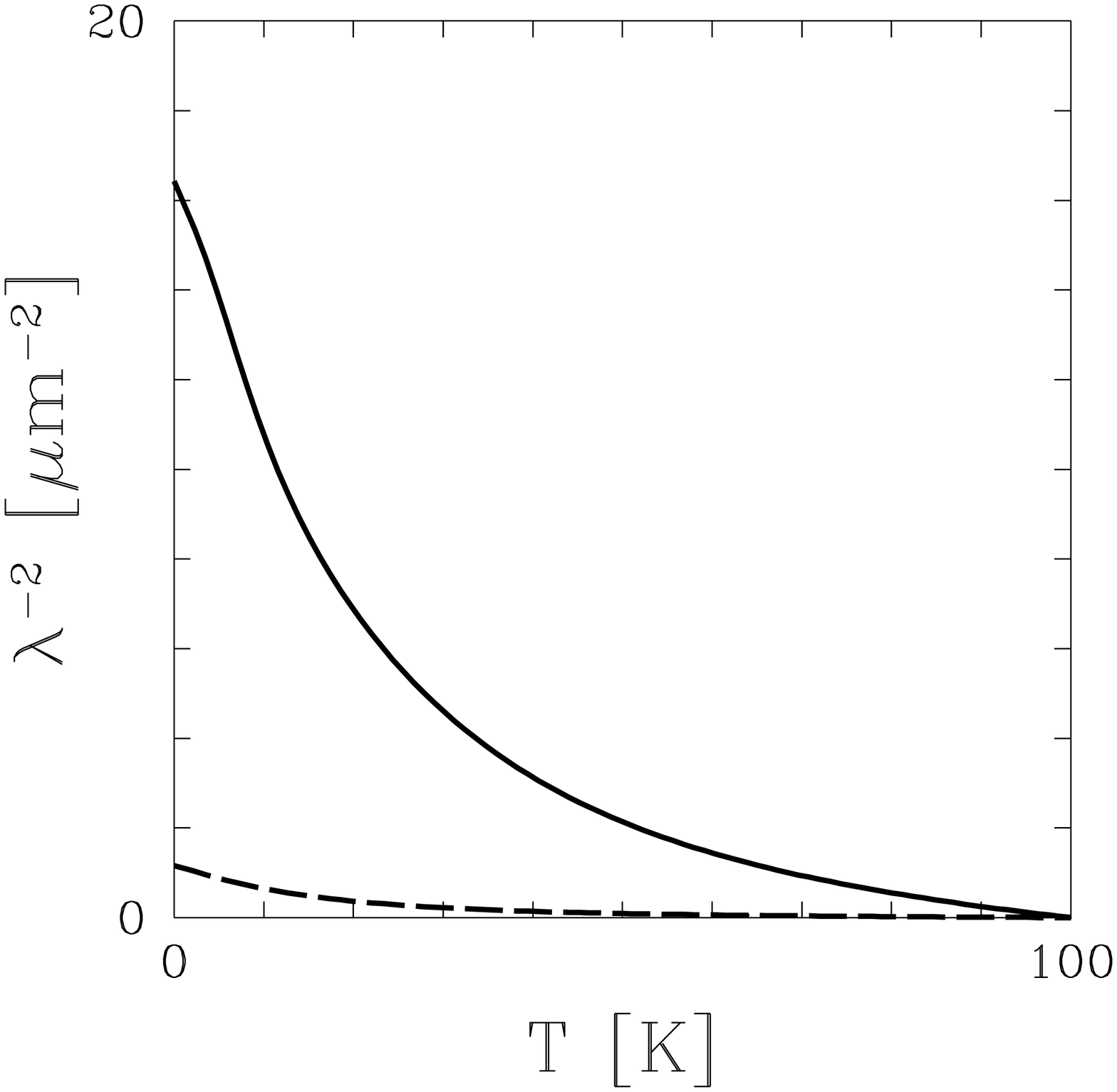} \\
    \makebox[\figwidtha][l]{\large (g)} &
    \makebox[\figwidthb][l]{\large (h)} &
    \makebox[\figwidthc][l]{\large (i)} \\
    \epsfxsize=\figwidtha \epsfbox{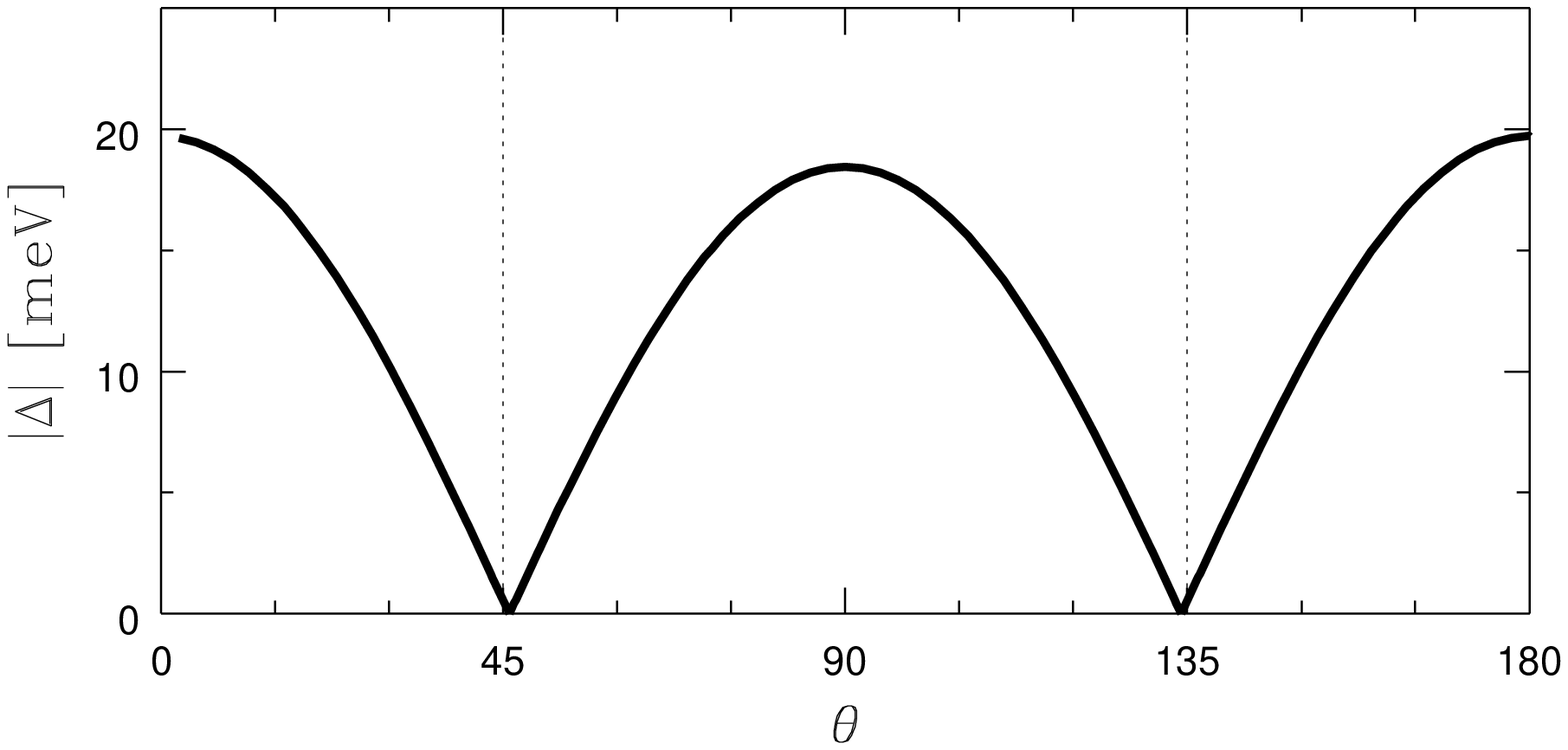} &
    \epsfxsize=\figwidthb \epsfbox{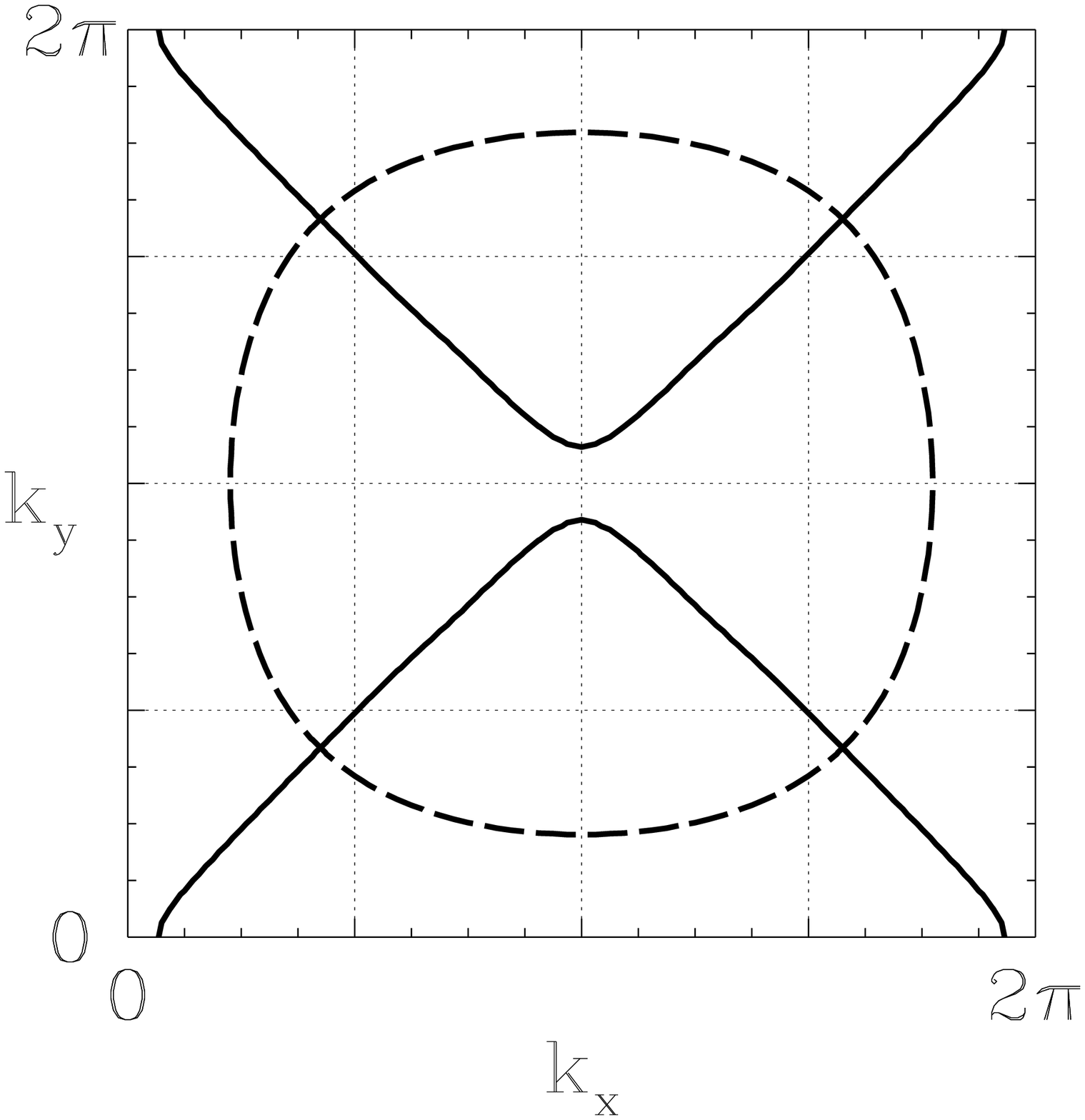} &
    \epsfxsize=\figwidthc \epsfbox{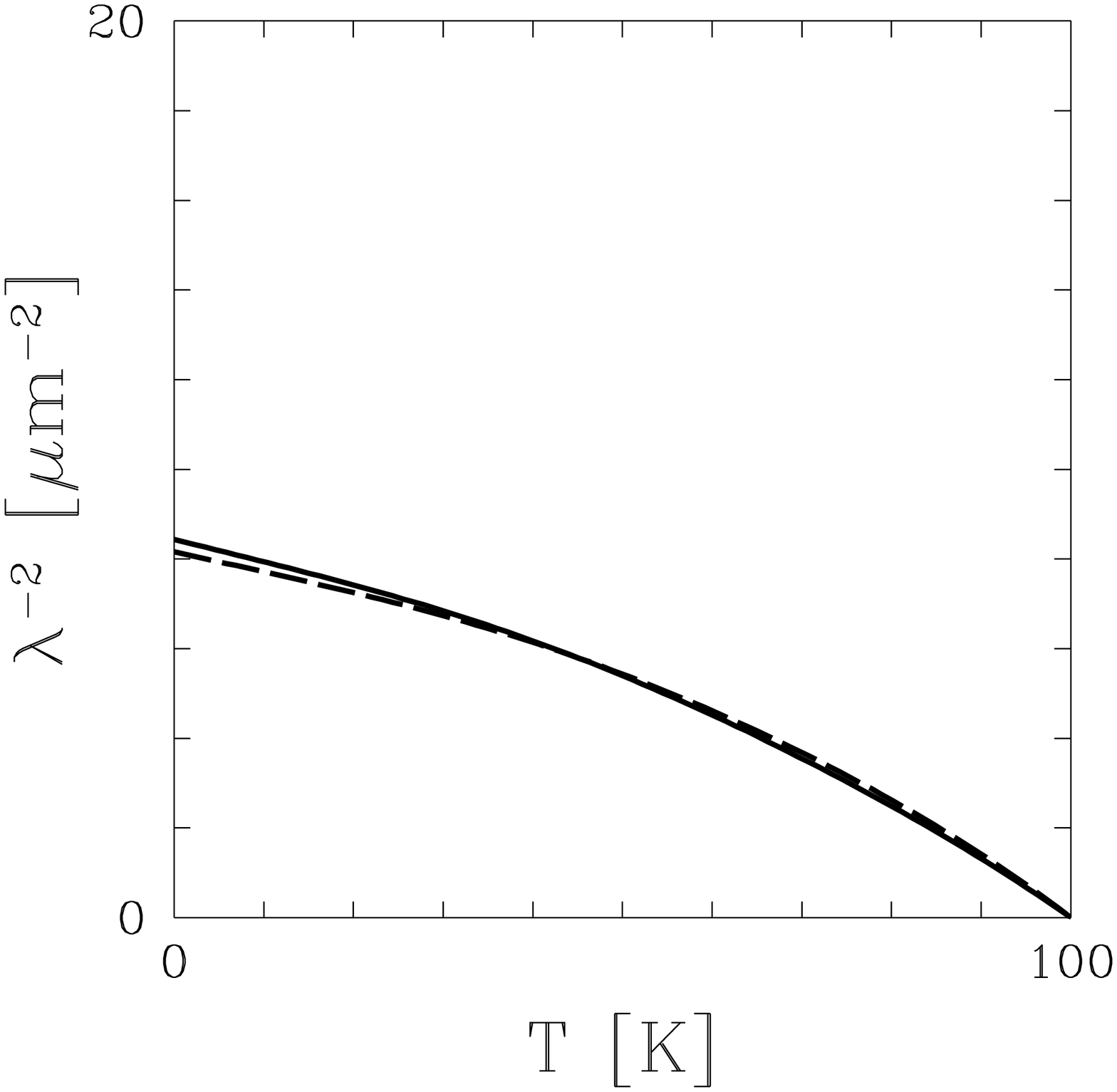}
  \end{tabular} \end{center}
\caption{
Results for a triplanar model with some off diagonal pairing on the
chains but with $g_{22}=0$, i.e. no direct pairing in the chains
themselves.  The chains become superconducting but with a small gap
amplitude.  Top frames apply to the odd (orthorhombic) CuO$_2$ band,
the middle three to the chain (CuO) band and the lower three frames to
the even (tetragonal) CuO$_2$ band.  Frames (a), (d) and (g) give the
absolute value of the gap as a function of angle $\theta$ along the
Fermi surface measured clockwise from the $k_x$ axis.  Note the lack
of tetragonal symmetry in the gap and the shift off 45$^\circ$ and
135$^\circ$ of the gap nodes.  Frame (b), (e) and (h) give the Fermi
surface (dashed curve) and gap zero contours (solid curve).  The odd
CuO$_2$ plane band is orthorhombic as are the chains while the even
CuO$_2$ band Fermi surface retains tetragonal symmetry.  Contributions
to the inverse square of the penetration depth as a function of
temperature are given in frames (c), (f) and (i) for $a$-direction
(dashed) and $b$-direction (solid).  The odd orthorhombic band
contributes almost twice as much to the $b$- as to the $a$-direction
while the even tetragonal band makes almost equal contributions.  The
chains contribute mainly in the $b$- direction. }
\label{fig8}
\end{figure*}

     As a second illustration of a triplanar model, we show in Fig.\ \ref{fig8}
results for model parameters chosen so that the chains also become
superconducting but by virtue of off diagonal coupling and not because
there is intrinsic pairing in the chains themselves.  That is $g_{22}$
is assumed to be zero.  Specifically, we assume:
\begin{equation}
g_{\alpha\beta}=
\left[\begin{array}{ccc}
44.4 & 10.0 & 0.0\\
10.0 & 0.0 & 10.0\\
0.0 & 10.0 & 44.4
\end{array}\right],
\label{interaction2.eq}
\end{equation}
\noindent
while the band parameters remain the same as in the previous cases.
In Fig.\ \ref{fig8}(b), (e) and (h), we show, respectively, odd, chain
and even band Fermi surface as the dashed curves and contours of gap
zeros is the solid curves.  The gap solution is:
\begin{equation}
\mit\Delta_{{\bf k},\alpha}=
\left[\begin{array}{ccc}
1.21 & -1.35 & 7.77\\
0.65 & -0.34 & 3.79\\
-0.59 & 0.15 & 11.17
\end{array}\right]
\left[\begin{array}{c}
\eta_{\bf k}^{(s_\circ)}\\
\eta_{\bf k}^{(s_x)}\\
\eta_{\bf k}^{(d)}
\end{array}\right].
\label{soln4.eq}
\end{equation}

The absolute values of the gaps as a function of angle $\theta$ along
odd, chain and even Fermi surface are shown in fig. 8(a), (d) and (g).
Because no pairing was assumed to act in the chains, the gap is
smallest in frame (d) with maximum $2\mit\Delta/k_{\rm B}T_{\rm c}$
equal to 1.88 as opposed to 4.58 and 2.98 for even and odd band,
respectively.  Frames (c), (f) and (l) show that odd orthorhombic band
makes a very different contribution to the $b$-direction (solid curve)
penetration depth than to the $a$-direction; the difference in zero
temperature value being of order 2.  The differences are even more
pronounced in the case of the chains which contribute very little to
the $a$-direction (dashed curve) while the even tetragonal Fermi
surface makes almost equal contribution to $a$- and $b$-directions.
Because of the small energy scale present on the chains, which can be
traced to the assumption $g_{22}=0$, i.e. zero pairing on the chains,
this model will not lead to good agreement with experiment.  To
achieve this, a final model is considered with the same single gap for
both CuO$_2$ planes and for the chains.  If we take all
$g_{\alpha\beta}={\rm constant} =18.4$, we obtain a single gap
\begin{equation}
\mit\Delta_{\bf k}=
\left[\begin{array}{ccc}
-0.363 & -0.502 & 10.14
\end{array}\right]
\left[\begin{array}{c}
\eta_{\bf k}^{(s_\circ)}\\
\eta_{\bf k}^{(s_x)}\\
\eta_{\bf k}^{(d)}
\end{array}\right].
\label{soln5.eq}
\end{equation}

\begin{figure*}
  \begin{center}\begin{tabular}{c c c}
    \makebox[\figwidtha][l]{\large (a)} &
    \makebox[\figwidthb][l]{\large (b)} &
    \makebox[\figwidthc][l]{\large (c)} \\
    \epsfxsize=\figwidtha \epsfbox{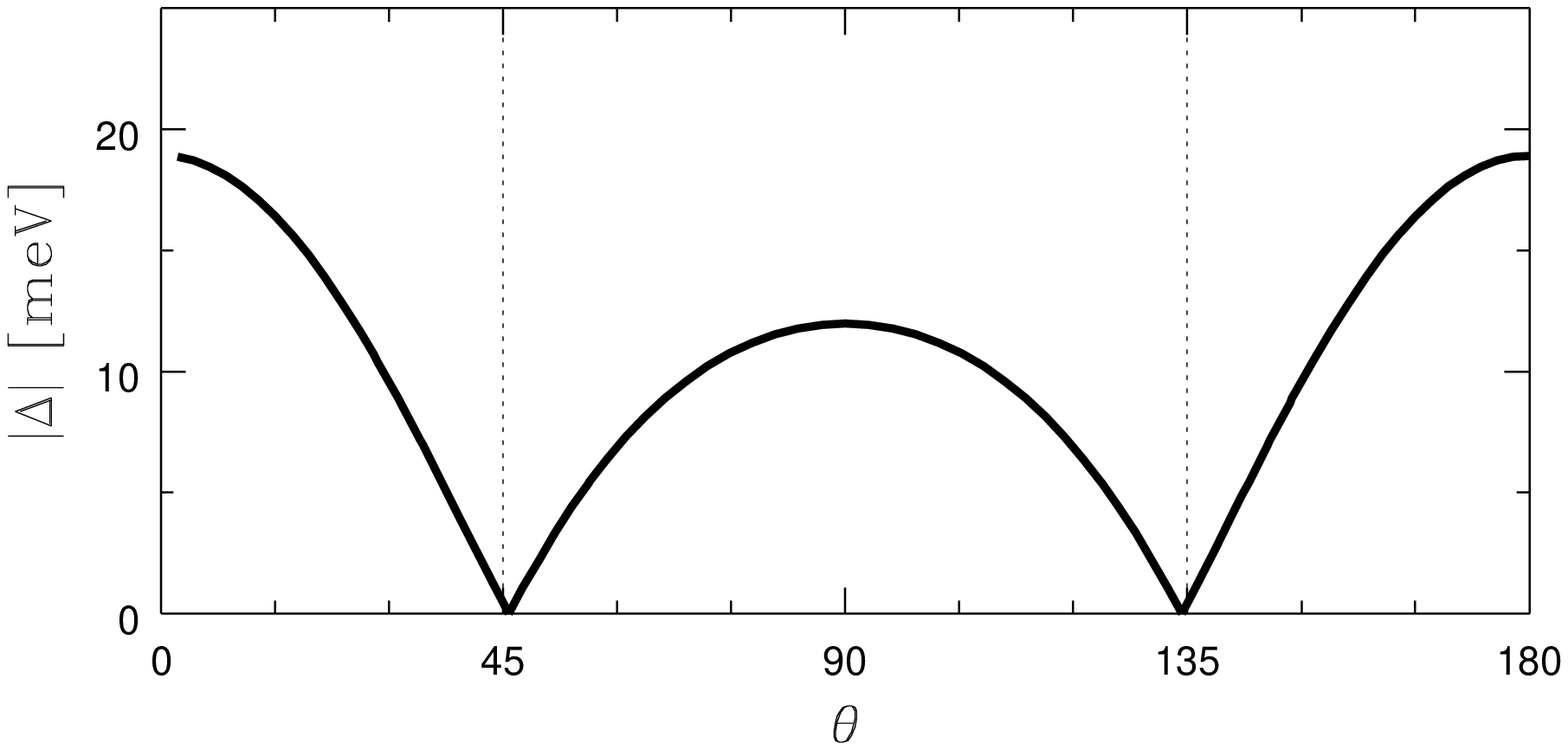} &
    \epsfxsize=\figwidthb \epsfbox{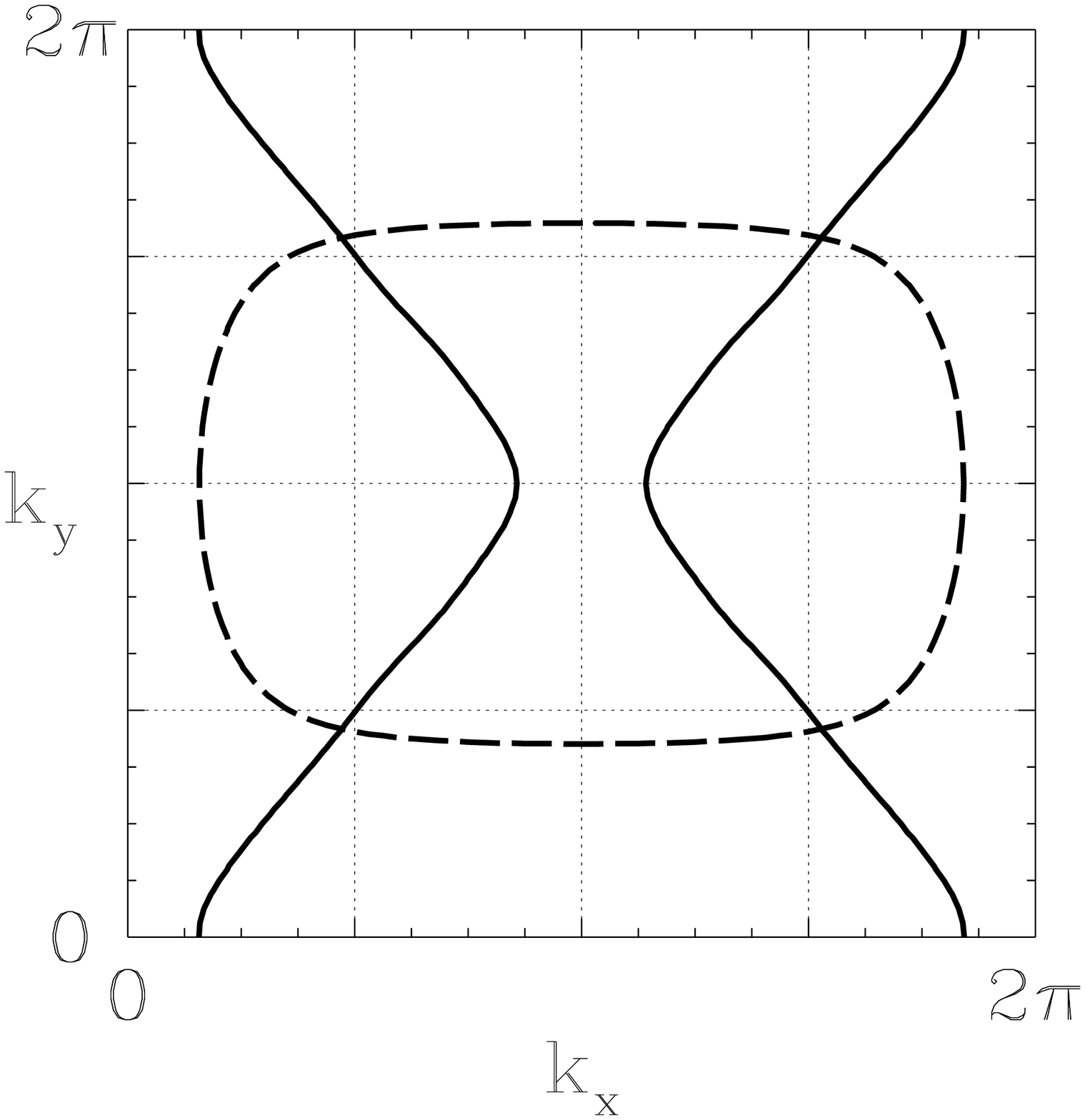} &
    \epsfxsize=\figwidthc \epsfbox{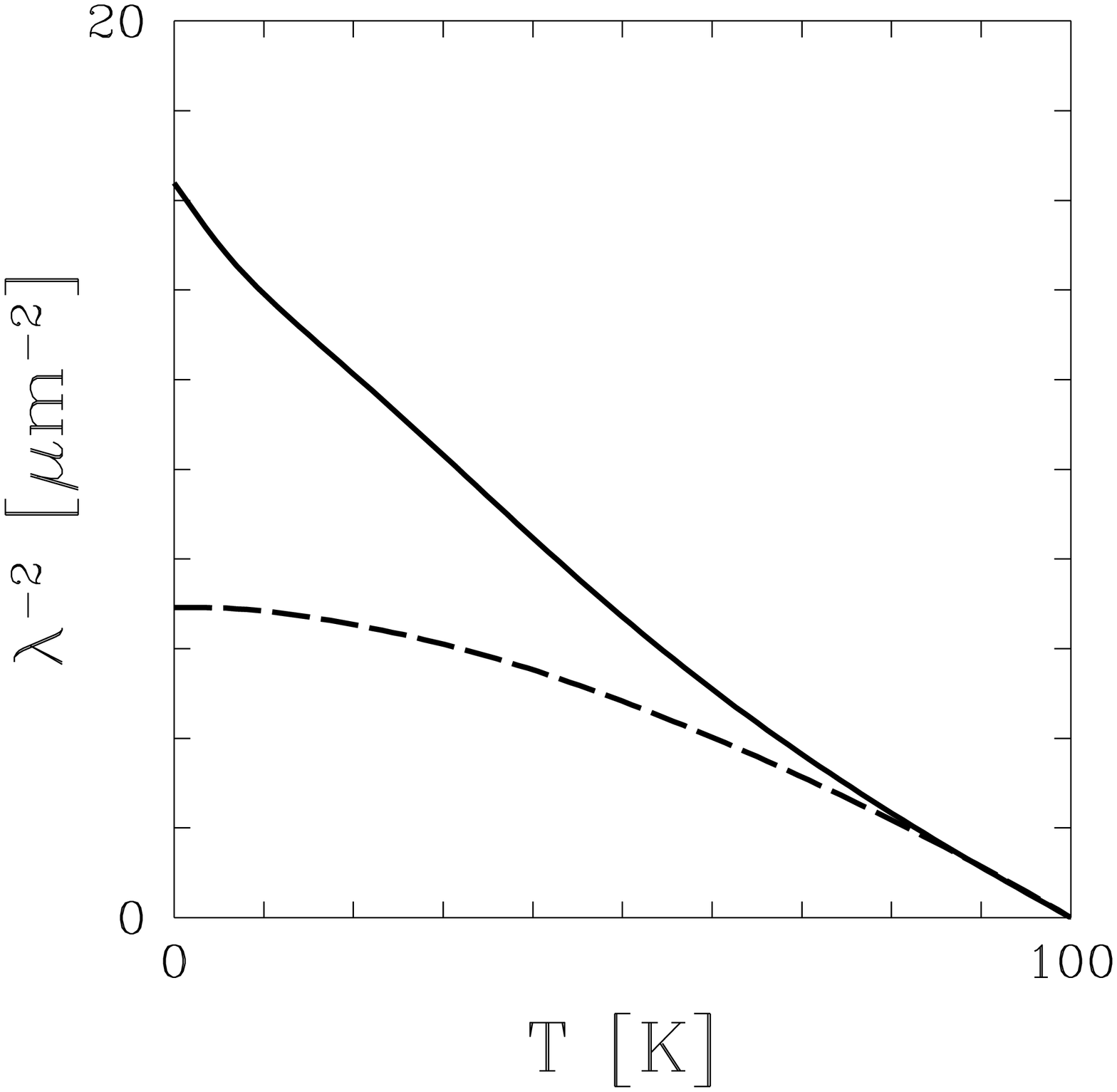} \\
    \makebox[\figwidtha][l]{\large (d)} &
    \makebox[\figwidthb][l]{\large (e)} &
    \makebox[\figwidthc][l]{\large (f)} \\
    \epsfxsize=\figwidtha \epsfbox{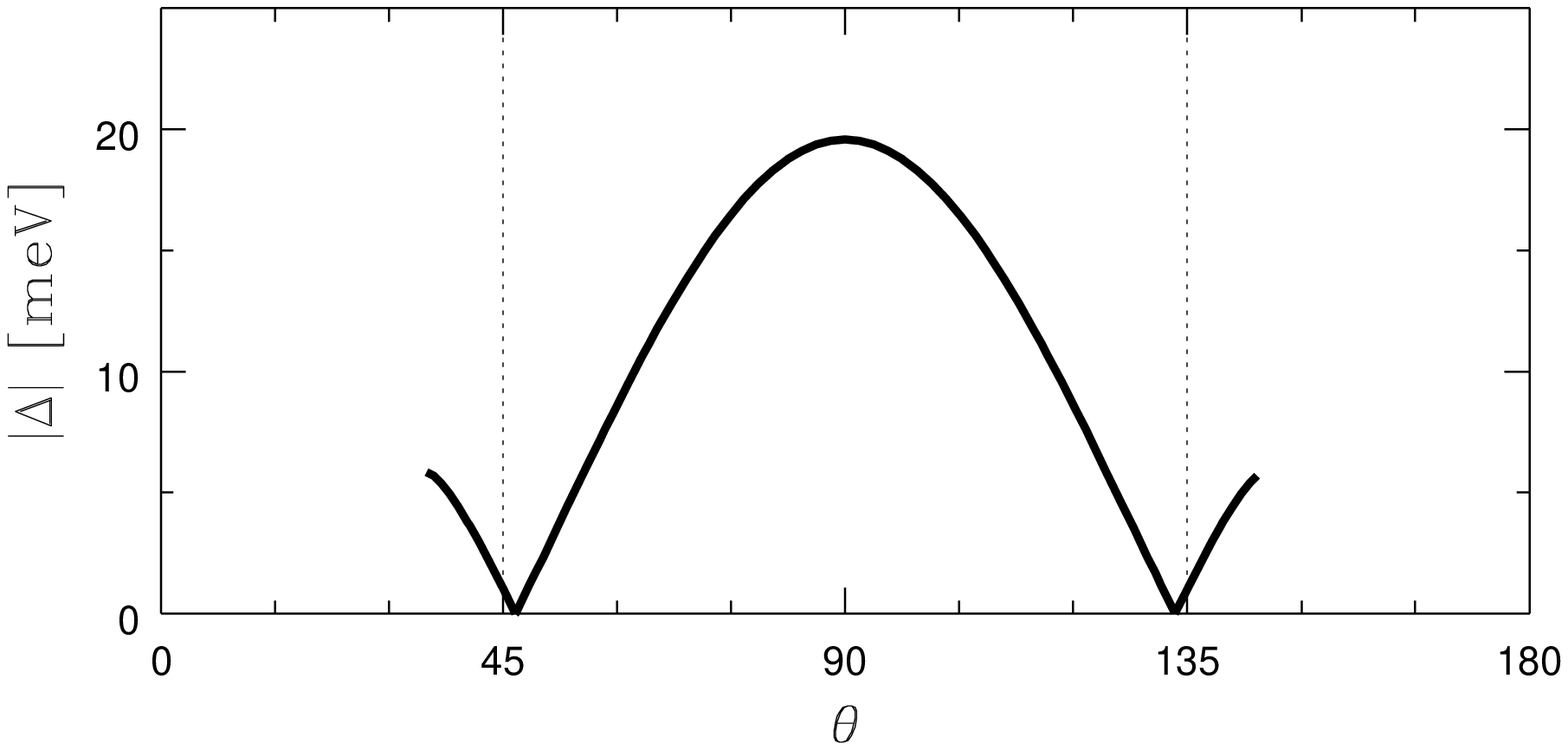} &
    \epsfxsize=\figwidthb \epsfbox{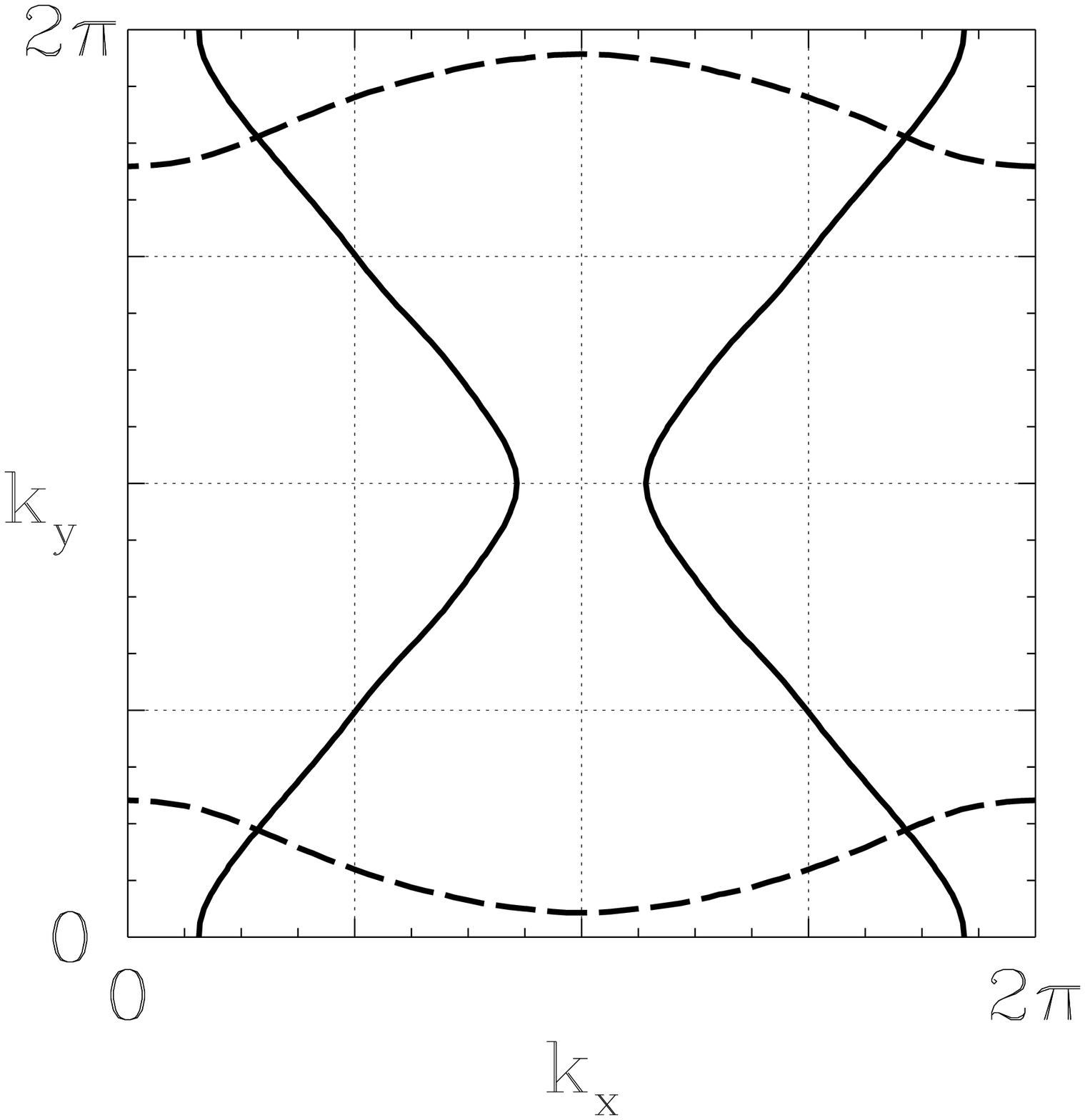} &
    \epsfxsize=\figwidthc \epsfbox{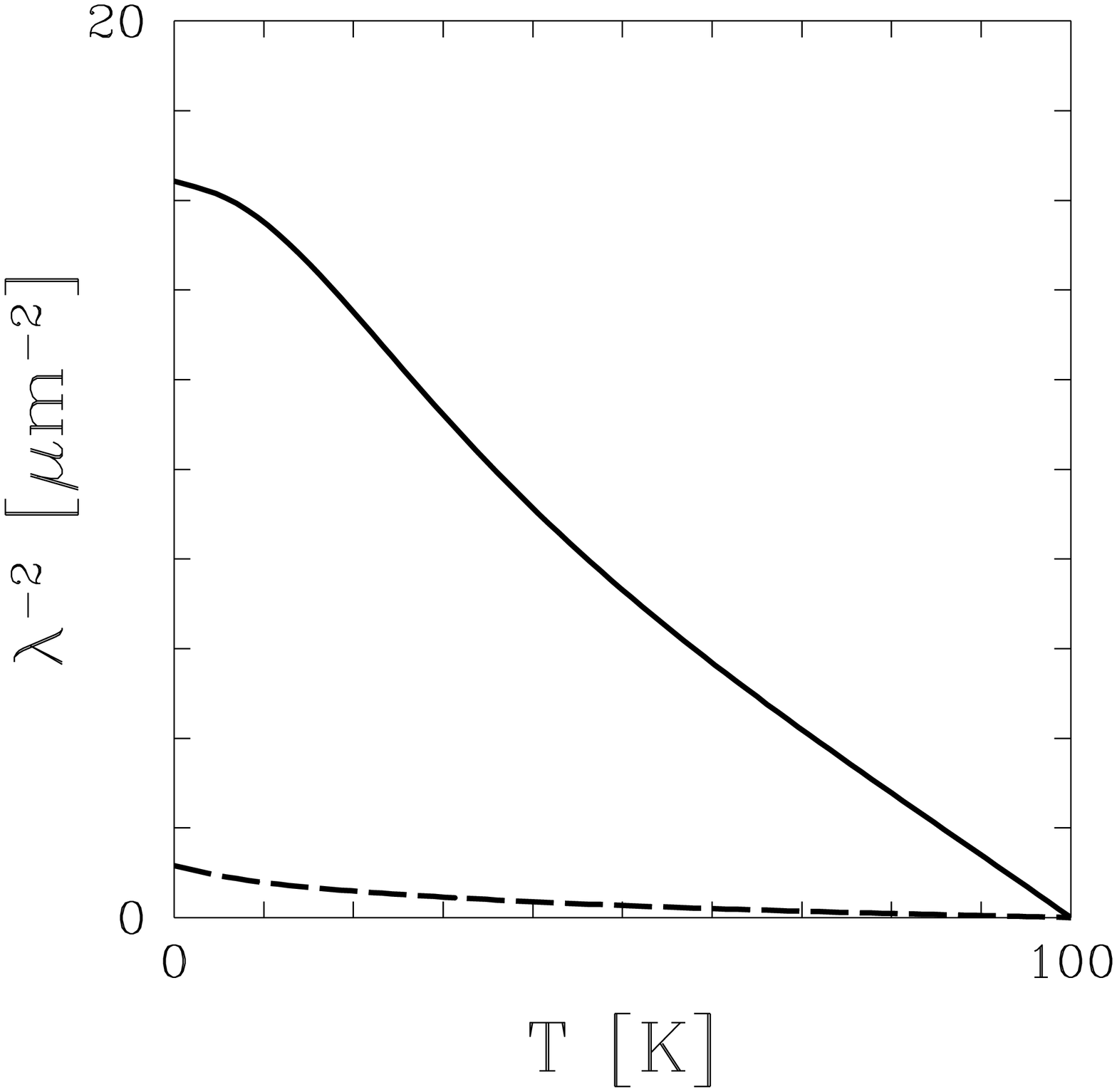} \\
    \makebox[\figwidtha][l]{\large (g)} &
    \makebox[\figwidthb][l]{\large (h)} &
    \makebox[\figwidthc][l]{\large (i)} \\
    \epsfxsize=\figwidtha \epsfbox{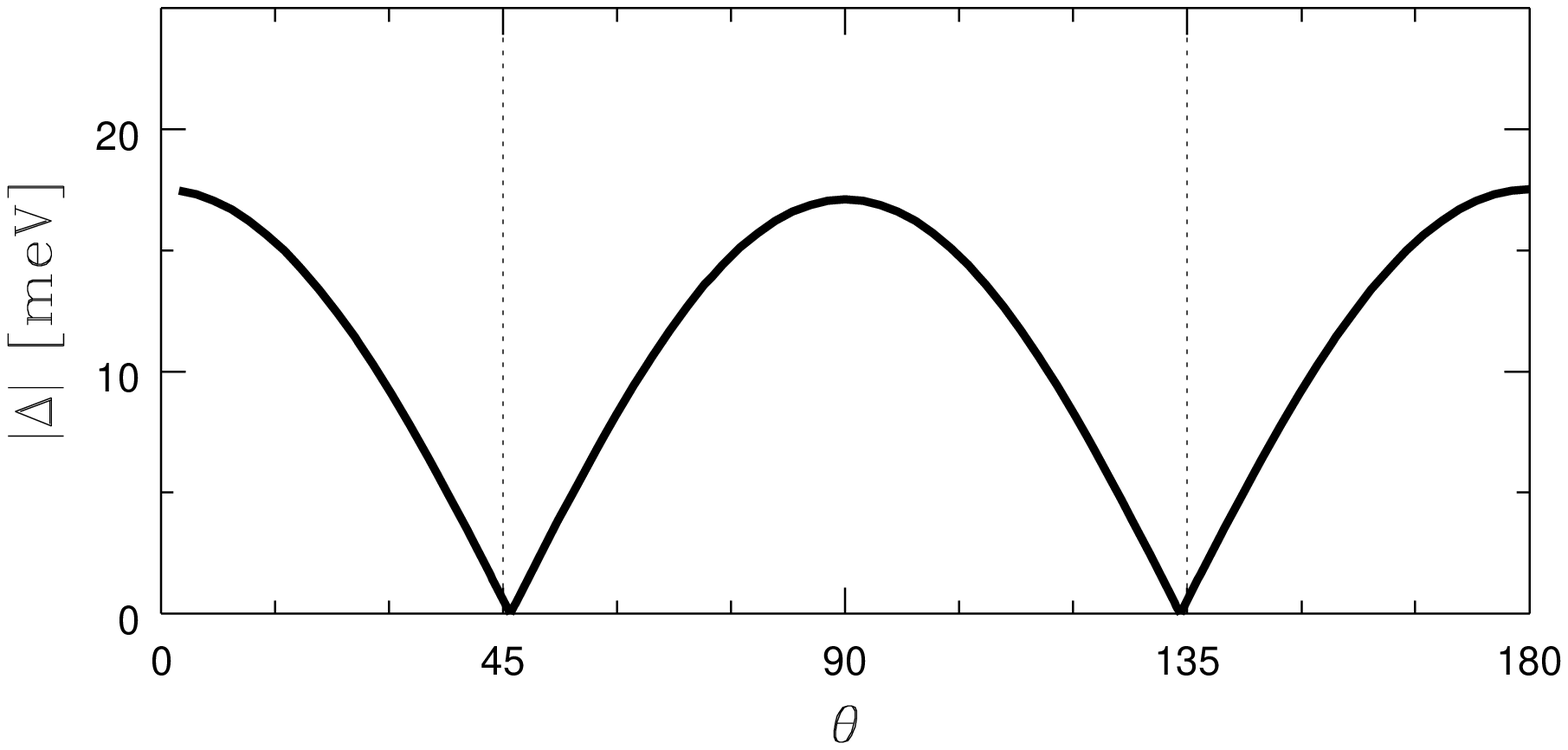} &
    \epsfxsize=\figwidthb \epsfbox{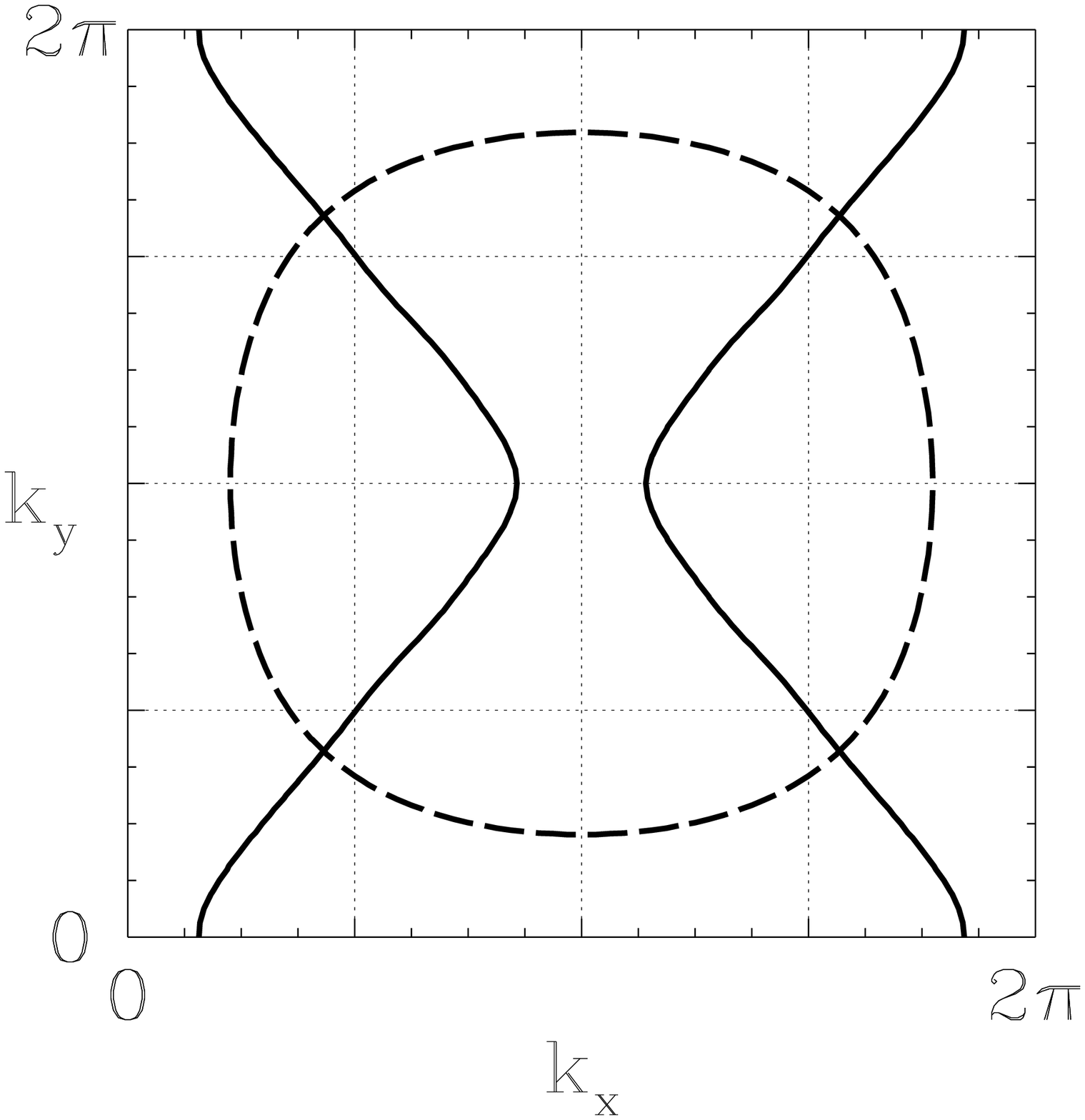} &
    \epsfxsize=\figwidthc \epsfbox{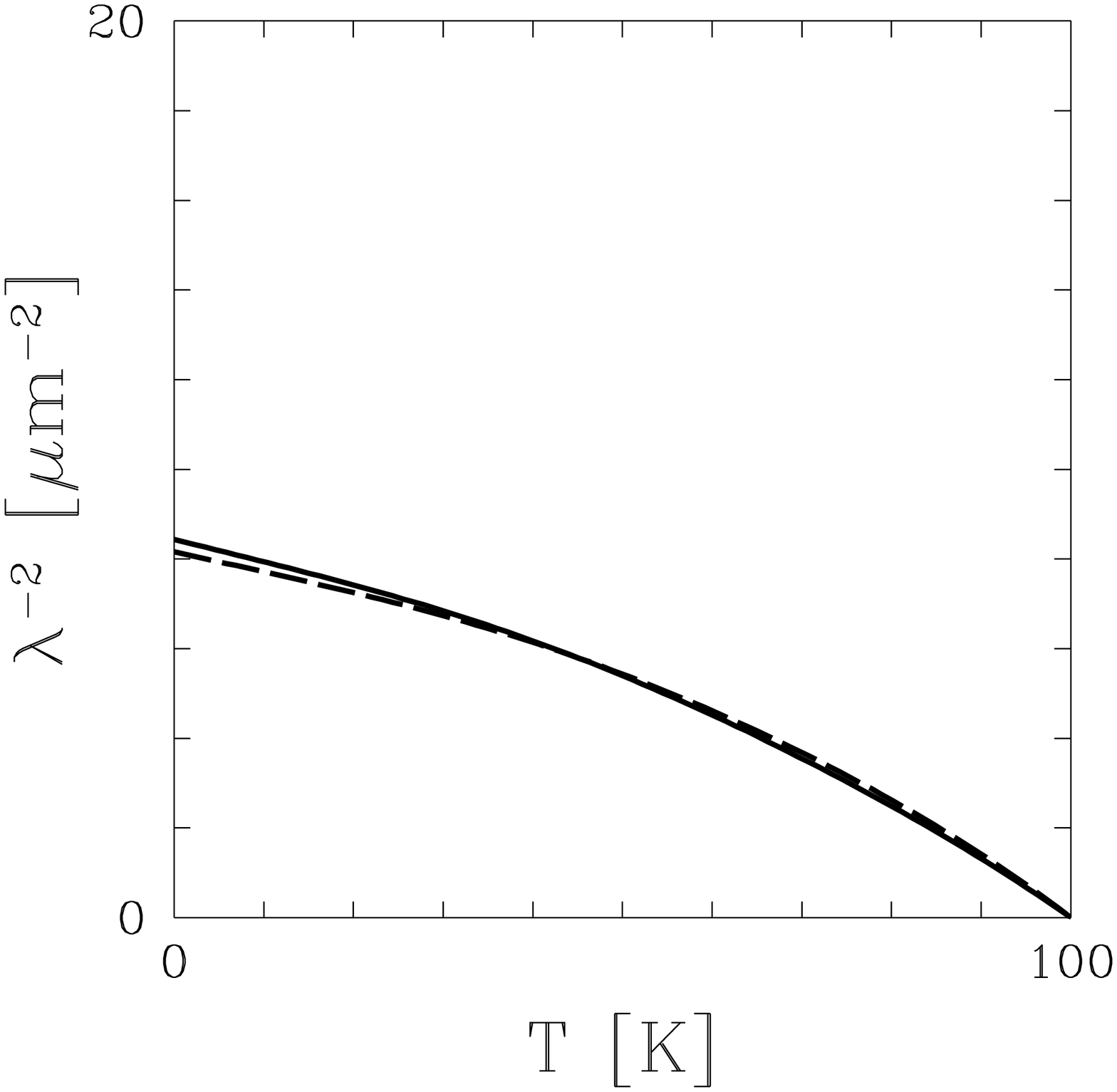}
  \end{tabular} \end{center}
\caption{
Results for a trilayer model with pairing interaction having the same
value on and off diagonal, i.e. $g_{\alpha\beta}={\rm constant}$ for
all $\alpha$ and $\beta$.  This assumption ensures that a single gap
applies to both the CuO$_2$ planes and the CuO chain.  The three upper
frames apply to the odd CuO$_2$ orthorhombic band, the middle three to
the CuO chain band and the lower three to the even CuO$_2$
(tetragonal) band.  Frames (a), (d) and (g) gives the gap on the Fermi
surface as a function of angle $\theta$ measured clockwise from the
$k_x$-axis.  Note the lack of tetragonal symmetry and the ``missing
parts'' in the chain case which reflects the geometry of the quasi one
dimensional Fermi surface.  Frame (b), (e) and (h) give the Fermi
contours (dashed line) for odd CuO$_2$ (orthorhombic) band, chain like
CuO band and even CuO$_2$ (tetragonal) band.  The solid curve are the
gap zero contours and are the same in all three figures.  Frames (c),
(f) and (i) give the contributions to the inverse square of the
penetration depth as a function of temperature for the $a$-direction
(dashed) and the $b$-direction (solid).  Note that the odd
orthorhombic band contributes almost twice as much to the $b$- as it
does to the $a$-direction in contrast to the even (tetragonal) band
which makes nearly equal contributions.  The chains contribute mainly
to the $b$-direction. In this model part of the $a$-$b$ asymmetry in
the penetration depth comes from the orthorhombic nature of the {\bf
even} plane Fermi surface of frame (b).}
\label{fig9}
\end{figure*}

\begin{figure}
  \postscript{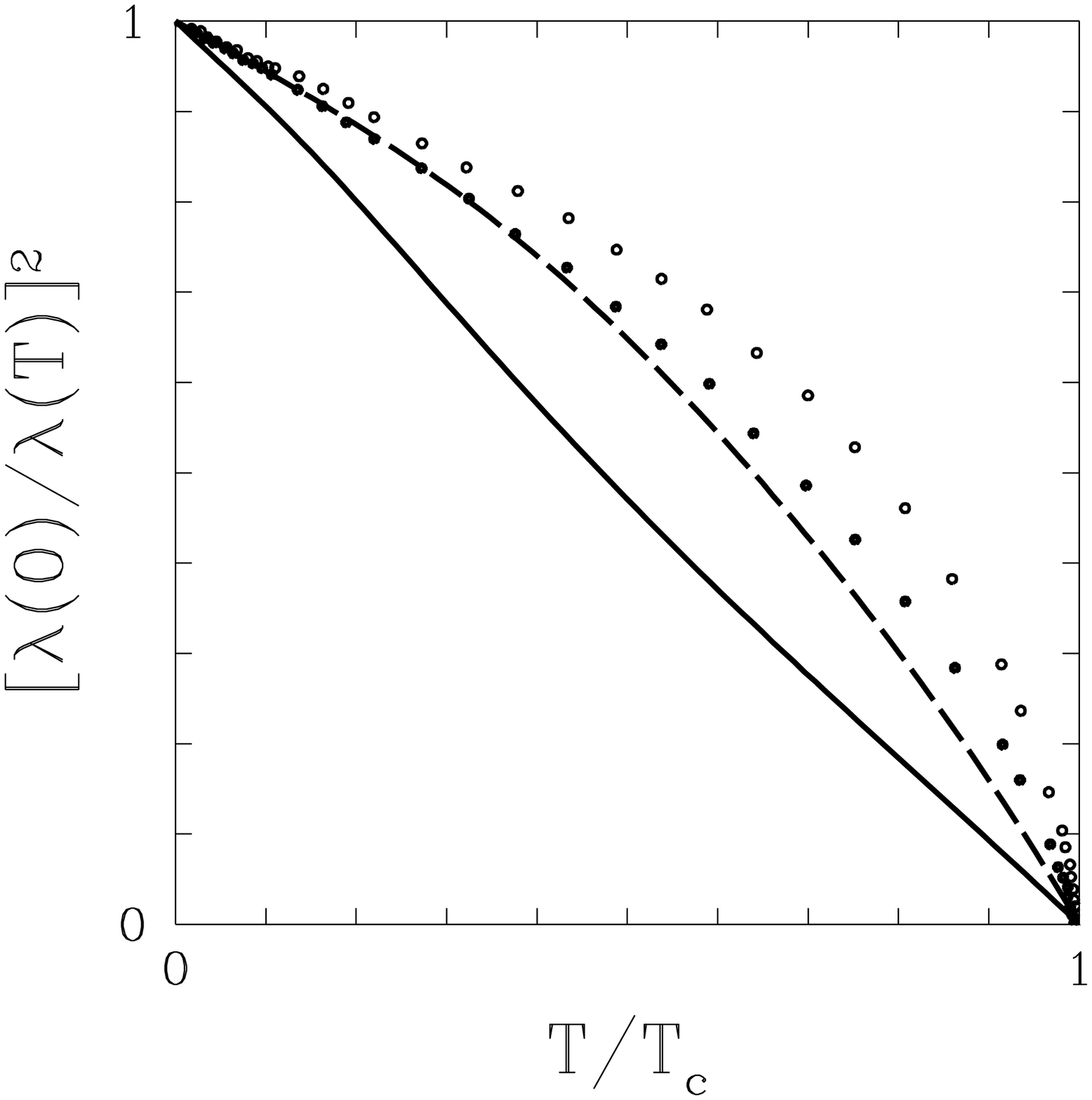}
\caption{
Total contributions to the inverse square of the penetration depth
normalized to its zero temperature value as a function of temperature.
The solid curve applies to the $b$-direction and dashed to the
$a$-direction.  The dots are experimental data\protect\cite{basov} with the
$b$-direction falling below $a$-direction.  The model is that of
Fig.\ \protect\ref{fig9} with pairing interaction assumed constant
(i.e. $g_{\alpha\beta}={\rm constant}$) independent of $\alpha$ and
$\beta$ which leads to a single gap for all three planes in the
primitive cell.  This does not imply that it has $d_{x^2-y^2}$
symmetry because the bands do not have tetragonal symmetry and the gap
contains an admixture of $s_\circ$ and $s_{x^2+y^2}$ although it is
still dominantly $d_{x^2-y^2}$.  }
\label{fig10}
\end{figure}

Detailed results are illustrated in the nine frames of Fig.\
\ref{fig9}.  In Fig.\ \ref{fig9} (b), (e) and (h), the gap zeros are
all the same (solid curves) yet because the Fermi contours (dashed
curves) are different, the gap as a function of angle along the Fermi
surfaces are different as shown in Fig.\ \ref{fig9} (a), (d) and (g)
for odd, chain and even Fermi surfaces, respectively.  The maxima in
$2\mit\Delta/k_{\rm B}T_{\rm c}$ are respectively 4.39, 4.55 and 4.07.
Resulting contributions to the $a$- (dashed) and $b$- (solid)
direction penetration depth are found in Fig.\ \ref{fig9} (c), (f) and
(i).  The odd orthorhombic Fermi surface again contributes about twice
as much to $b$- and to $a$- direction at low temperature while the
even one contributes equally.  The chains mainly contribute to the
$b$-direction.  The sum of all contributions are shown in Fig.\ \ref{fig10} and
compared with experiment.  The agreement is satisfactory with
$a$-direction curve (dashed) above the $b$-direction (solid) as
observed.  The ratio $\lambda_a/\lambda_b$ was $\sim 1.6$ in the
calculations, close to the observed value.

\section{Conclusions}

     We have presented results of calculations for the temperature
dependence of the in plane penetration depth in a trilayer model of
YBa$_2$Cu$_3$O$_7$ which includes two CuO$_2$ planes and a set of CuO
chains oriented along the $b$-direction.  Transverse hopping matrix
elements couple the three planes together leading to three coupled
bands in which the gap parameters is an admixture of $d_{x^2-y^2}$,
$s_{x^2+y^2}$ and $s_\circ$ symmetries.  While the $s$ and $d$
symmetry types belong to two different irreducible representations of
the tetragonal crystal point group, they belong to the same
representation in the orthorhombic case which is the case we are
dealing with here because of the presence of the chains.
Nevertheless, it is found that the penetration depth remains linear in
temperature at low temperature as a result of nodes in the gap
crossing the Fermi contours.  To compare with experiment, the ratio of
$\lambda_a/\lambda_b$ was kept near the observed value in all our
calculations.  With this constraint, it was found that the observed
normalized temperature variation of the penetration depth in $a$- and
$b$-directions is difficult to understand in models where no
intraplane pairing is introduced in the chains even if a large amount
of interplane pairing is considered.  The most natural explanation of
the data is obtained when the pairing interaction is assumed not to
depend significantly on sub-band and is taken to be independent of
$\alpha$ and $\beta$, i.e. $V_{{\bf k},{\bf k}^\prime,
\alpha\beta}\equiv V_{{\bf k},{\bf k}^\prime}$. Also the zero
temperature anisotropy in $\lambda_{a,b}$ can be traced to the
orthrhombicity of the band structure which resides in the chains and
hybridized planes.

\section*{Acknowledgement}

     Work supported in part by the Canadian Institute for Advanced
Research ({\sc ciar}) and the Natural Sciences and Engineering
Research Council of Canada ({\sc nserc}).  We have benefited from
conversations with W. Atkinson.

\begin{references}

\bibitem[\dagger]{me}
Electronic address: {\tt odonovan@mcmaster.ca}

\bibitem{basov}
D.N. Basov, R. Liang, D.A. Bonn, W.N. Hardy, B. Dabrowski,
M. Quijada, D.B. Tanner, J.P. Rice, D.M. Ginsberg and T. Timusk
Phys.\ Rev.\ Lett.\ {\bf 74}, 598 (1995).

\bibitem{zhang}
K. Zhang, D.A. Bonn, S. Kamal, R. Liang, D.J. Baar, W.N. Hardy,
D. Basov and T. Timusk, Phys.\ Rev.\ Lett.\ {\bf 73}, 2484 (1994).

\bibitem{bonn}
D.A. Bonn, S. Kamal, K. Zhang, R. Liang and W.N. Hardy, J.\ Phys.\
Chem.\ Solids {\bf 56}, 1941 (1995).

\bibitem{friedmann}
T.A. Friedmann, M.W. Fabin, J. Giapintzakis, J.P. Rice and
D.M. Ginsberg, Phys.\ Rev.\ {\bf B42}, 6217 (1990).

\bibitem{iye}
Y. Iye in {\it Physical Properties of High Temperature
Superconductors} Vol III, p. 285 (edited by D.M. Ginsberg, World
Scientific, Singapore (1992).

\bibitem{gagnon}
R. Gagnon, C. Jupien and L. Taillefer, Phys.\ Rev.\ {\bf B50}, 3458
(1994).

\bibitem{takenaka}
K. Takenaka, K. Mizuhashi, H. Takage and S. Uchida, Phys.\ Rev.\ {\bf
B50}, 6534 (1994).

\bibitem{atkinson}
W.A. Atkinson and J.P. Carbotte, Phys. Rev. {\bf B 51}, 1161 (1995);
Phys. Rev. {\bf B 51}, 16371 (1995).

\bibitem{atkinson2}
W.A. Atkinson and J.P. Carbotte, Phys. Rev. {\bf B 52}, 6894 (1995);
Phys. Rev. {\bf B 52}, 10601 (1995).

\bibitem{odonovan10}
C. O'Donovan and J.P. Carbotte, to appear in Phys.\ Rev.\ {\bf B 54}, (1996).

\bibitem{mmp}
A.J. Millis, H. Monien and D. Pines, Phys.\ Rev.\ {\bf B 42}, 167
(1990).

\bibitem{odonovan4}
C. O'Donovan and J.P. Carbotte, Phys.\ Rev.\ {\bf B 52}, 4548 (1995).

\bibitem{odonovan3}
C. O'Donovan, D. Branch, J.P. Carbotte and J. Preston, Phys. Rev. {\bf
B 51}, 6588 (1995).

\bibitem{odonovan1}
C. O'Donovan and J.P. Carbotte, Physica {\bf C} 252, 87 (1995).



\end{references}
\end{document}